\documentclass{aa}  
\usepackage{graphicx}
\usepackage{txfonts}
\usepackage{lscape}         
\begin{document}
%
%

   \title{Optical properties of silicon carbide for astrophysical applications}

   \subtitle{I. New laboratory infrared reflectance spectra and optical constants}

   \author{K. M. Pitman
             \inst{1, 2},
	     A. M. Hofmeister\inst{2},
	     A. B. Corman\inst{3}
	     \and
	     A. K. Speck\inst{3}
          }

   \offprints{K. M. Pitman}

   \institute{Jet Propulsion Laboratory, California Institute of 
             Technology, Pasadena, CA 91109, USA\\
             \email{Karly.M.Pitman@jpl.nasa.gov}
       \and
	   Department of Earth and Planetary Sciences, Washington
	   University, St. Louis, MO 63130, USA\\				
             \email{hofmeist@levee.wustl.edu}
      \and
	   Department of Physics and Astronomy, University of
	   Missouri-Columbia, Columbia, MO 65211, USA\\
	   \email{abcp68@mizzou.edu, speckan@missouri.edu}
}	

\date{Received 16 August 2007; accepted 01 March 2008}

 
  \abstract
  {} 
  {The SiC optical constants are fundamental inputs for radiative transfer 
(RT) models of astrophysical dust environments.  However, previously 
published values contain errors and do not adequately represent the 
bulk physical properties of the cubic ($\beta$) SiC polytype usually 
found around carbon stars.  We provide new, uncompromised optical 
constants for $\beta$- and $\alpha$-SiC derived from single-crystal 
reflectance spectra 
\thanks{Tables 3--7 are only available in electronic form at the CDS
via anonymous ftp to cdsarc.u-strasbg.fr (130.79.128.5)
or via http://cdsweb.u-strasbg.fr/cgi-bin/qcat?J/A+A/}
and investigate quantitatively (i) whether
there is any difference between 
$\alpha$- and $\beta$-SiC that can be seen in infrared (IR) 
spectra and optical functions and
(ii) whether weak features from $\lambda$~$\sim$~12.5--13.0~$\mu$m need to be fitted.}
 {We measured mid- and far-IR 
 reflectance spectra for two samples of 3C ($\beta$-)SiC and four samples of 
6H ($\alpha$-)SiC.  For the latter group, we acquired polarized data 
($E \bot c$, $E \| c$ orientations).  We calculated the 
real and imaginary parts of the complex refractive index 
($n(\lambda) + ik(\lambda)$) and the ideal absorption coefficients via 
classical dispersion fits to our reflectance spectra.} 
 {We find that $\beta$-SiC and $\vec{E} \bot \vec{c}$ $\alpha$-SiC have 
almost identical optical functions but that 
$n(\lambda)$ and $k(\lambda)$ for $\vec{E} \| \vec{c}$ $\alpha$-SiC 
are shifted to lower frequency.  Peak positions determined for both 
3C ($\beta$-) and 6H ($\alpha$-)SiC polytypes agree with Raman 
measurements and show that a systematic error of 4 cm$^{-1}$ exists 
in previously published IR analyses, attributable to
inadequate resolution of older instruments for the steep, sharp modes of SiC.
Weak modes are present for samples with impurities.  Our 
calculated absorption coefficients are much higher than laboratory 
measurements.  Whereas astrophysical dust grain sizes remain fairly 
unconstrained, SiC grains larger than about 1~$\mu$m in diameter 
will be opaque at frequencies near the peak center.}  
 {Previous optical constants for SiC do not reflect the true bulk 
properties, and they are only valid for a narrow grain size range.  The 
new optical constants presented here will allow narrow constraints to 
be placed on the grain size and shape distribution that dominate in 
astrophysical environments.}
   

   \authorrunning{Pitman et al.}
   \titlerunning{Optical properties of SiC}

   \maketitle
%

\section{Introduction}

Dust grains play an essential role in star formation, contribute to 
interstellar processes, and are important to radiatively-driven 
mass loss from evolved stars.  Thus, a detailed understanding of cosmic dust 
is necessary to determine its role in many astrophysical 
environments.  Laboratory studies of space-borne dust have 
furthered our understanding of the physical properties
(e.g., composition, size, shape, crystal structure, and porosity) and,
in turn, the origin of these grains.
Silicon carbide (SiC) were the first presolar dust grains found 
(Bernatowicz et al. 1987) and are thus extremely important.
It is inferred that $\sim$\,99\% of meteoritic SiC
grains were formed around carbon (C-)stars for which the
carbon to oxygen ratio (C/O) is greater than unity 
(e.g., Bernatowicz et al. 2006).  
These pristine samples of stardust have provided details 
of the grain size and crystal structure of circumstellar dust. 
The most significant result from presolar SiC grain studies is that 
nearly all ($\sim$80\%) are of the 
$\beta$-polytype (i.e., have a cubic crystal structure; see \S~2) and 
the 6H-$\alpha$-polytype is never found.
The formation of SiC in C-rich environments was predicted nearly 40
years ago via equilibrium condensation models (Friedemann 1969; Gilman 1969).
Subsequently, a broad feature located at $\sim$\,11\,$\mu$m 
similar to that found in laboratory infrared (IR) spectra of SiC grains was
discovered in the spectra of C-stars (Hackwell 1972; Treffers \& Cohen
1974).  The $\sim$\,11\,$\mu$m spectral feature is almost
ubiquitous amongst observed spectra of C-stars, although its precise
position, strength, and shape varies from star to star.  Whereas amorphous or
graphitic carbon grains dominate the circumstellar shells of C-stars, these 
species do not
have diagnostic IR features and merely contribute to the dust continuum.  
Consequently, the $\sim$\,11\,$\mu$m SiC feature has been used
extensively to infer physical properties of and processes occurring within
these circumstellar shells.  In addition, the lack of observational
evidence for the $\sim$\,11\,$\mu$m feature in spectral studies of the
interstellar medium (ISM) has placed an upper limit on the interstellar SiC
abundance.

Studies that attempt to quantify the abundance of SiC from the strength of its
$\sim$\,11\,$\mu$m feature require 
laboratory data in two forms: IR absorption measurements for direct 
comparison to observational spectra; and complex refractive indices 
derived from reflectance measurements and supplied to radiative transfer 
(RT) codes to assess the relative contributions of different minerals to 
a given observed spectrum (cf. Thompson et al. 2006).  The complex 
refractive indices of SiC used in astronomical studies derive primarily 
from three sources: Bohren \& Huffman (1983), P\'{e}gouri\'{e} (1988), 
and Laor \& Draine (1993).  However, there are several 
problems with these datasets. 

\begin{enumerate}

\item{Most complex refractive indices (optical functions, or 
``constants,'' $n(\lambda)$ and $k(\lambda)$) were calculated from reflectivity 
measurements of $\alpha$-SiC, rather than the dominant $\beta-$SiC.
In addition, the existing $\alpha$-SiC laboratory reflectance data 
were gathered at resolutions lower by at least a factor of two than can 
be attained with modern instrumentation.}

\item{Published works on SiC (e.g., Dorschner et al. 1977, Friedemann et al. 
1981,  Borghesi et al. 1985, Orofino et al. 1991, Papoular et al. 1998,
Mutschke et al. 1999, Andersen et al. 1999b and references therein) 
are predominantly laboratory absorbance studies of powder, not bulk, 
samples.  Thin films created from powders are imperfect and 
dispersions of powders tend to clump such that grains cluster, 
causing strong differences in the optical properties of the sample 
(cf., Huffman 1988).  The matrix in which the powder is dispersed 
influences the shape and profile of the bands, therefore, measurements 
made using this technique are not quantitative 
(Mutschke et al. 1999).  Measuring reflectances at different angles or a 
combination of reflectance and transmission of a 
single-crystal is optimal for calculating optical constants.}

\item{Whereas samples
from laser pyrolysis methods are considered to produce 
SiC particles most closely resembling those found in stellar environments
(c.f. Willacy \& Cherchneff 1998; Mutschke et al. 1999), such samples are
not appropriate for determining bulk (grain size independent) optical constants.
This is because laser pyrolysis methods do not produce crystals sufficiently
large for the collection of polarized IR reflectivity data (of
the order of several mm in diameter, cf. Hofmeister et al. 2003).}

\item{The \emph{KBr matrix correction\/} (cf. Friedemann et al. 1981)
has been applied to some laboratory
spectra of sub-micron diameter SiC grains dispersed in single-crystal
matrices.  
Studies of thin films and isolated nanoparticles of $\beta$-SiC 
have shown that this wavelength shift is unnecessary when measurements 
are made carefully (Speck et al. 1999; Cl\'{e}ment et al. 2003).  
Use of the KBr matrix correction caused a discrepancy between the IR spectra of
SiC polytypes found in meteorites and around C-stars, which has since
been resolved (see Speck et al. 1999).  However, the most widely used SiC
optical constant datasets predate this discovery and thus the 
KBr matrix correction was adopted by 
Bohren \& Huffman  (1983, p. 342) for spherical grains
 and P\'{e}gouri\'{e} (1988) and is indirectly included in the Laor \& Draine
(1993) dataset, which used both the P\'{e}gouri\'{e} (1988) and
Bohren \& Huffman (1983) data.}

\item{In addition, the most widely used SiC optical constants have been 
calculated
from compiled laboratory spectra which originated from different samples (cf.,
corundum; Speck et al. 2000).
For $\beta$-SiC, refractive indices have been compiled from various sources 
by Adachi (1999) from 0.13\,$\mu$m $\leq \lambda \leq$ 124\,$\mu$m, but 
modelers must extrapolate for the complex index of refraction in subregions 
along that interval (e.g., 0.5\,$\mu$m $\leq \lambda \leq$ 0.65\,$\mu$m; see 
Jiang et al. 2005).}

\item{The shortcomings of these datasets compromise the results of RT studies.
Martin \& Rogers (1987) Lorenz-Martins \& Lefevre (1993,
1994), Lorenz-Martins et al. (2001), Groenewegen (1995), Groenewegen et al.
(1998), Griffin (1990, 1993), Bagnulo et al. (1995, 1997, 1998) and others
have used these 
optical constants in RT models of C-stars in order to place limits on 
the SiC abundance.  Given the problems 
described above, the abundances derived in those studies are
probably incorrect.  These optical constants have also been used in modeling
environments that might include SiC (e.g., LMC stars, Speck et al. 2006;
PPNe, Clube \& Gledhill 2004; PNe, Hoare 1990). The veracity of those results
is now suspect.
Moreover, the existing SiC optical constants are valid only for a
limited grain size distribution.
The Borghesi et al. (1985) laboratory data, used by P\'{e}gouri\'{e} (1988) 
and consequently Laor \& Draine (1993), are based on a narrow grain size 
distribution ($a^{-2.1}$, with typical size of $a =$\,0.04\,$\mu$m), 
which is narrower than the modeled grain size distributions assumed by 
Laor \& Draine (1993); thus, these data are inappropriate for
analysing grain size effects (e.g., Bagnulo et al. 1995; see 
Speck et al. 2005).  Furthermore, these optical constants have been used 
to determine the maximum abundance of SiC grains in 21\,$\mu$m PPNe 
(Jiang et al. 2005).  These derived upper limits cannot be trusted.}

\item{Optical properties of SiC are also
used in studies of dust formation (e.g. Kozasa et al.
1996), hydrodynamics of circumstellar shells (e.g. Windsteig et al. 1997;
Steffen et al. 1997) and mean opacities (Ferguson et al. 2005; Alexander \&
Ferguson 1994).  Such studies are compromised.}

\item{The different datasets of SiC optical constants
produce different results.  For example, using a band strength 
based on the data presented by Bohren \& Huffman (1983) and 
Borghesi et al. (1985), Whittet et al. (1990) determined an
upper limit for the abundance of SiC in the ISM such that less than 5\%
of the available Si atoms could reside in SiC grains.  A similar study used
data from Borghesi et al. (1985) to constrain further the SiC abundance to
less than 4\% of the available Si atoms that could reside in SiC grains
(Chiar \& Tielens 2006).  However, Min et al. (2007) used the optical
constants derived by Laor \& Draine (1993) and found that 3\% of the
interstellar grains are SiC, implying that 9--12\% of the available
Si atoms are bound in SiC grains.  Clearly the precise choice of optical
constants and IR laboratory data has important consequences for
quantitative analyses of the impact of SiC grains on astrophysical
environments.}

\end{enumerate}

Given these issues, there is a need for new, more robust optical 
constants.  We address these problems by providing new 
$n$ and $k$ values obtained from high-resolution laboratory reflectance 
measurements of both $\alpha$- and $\beta$-SiC single-crystals samples. 
In \S~2--4, we present and interpret measurements of mid- and far-IR room
temperature reflectance spectra for several polytypes and orientations
($\vec{E} \bot c$, $\vec{E} \| c$) of commercially manufactured SiC: 
semiconductor grade purity 3C ($\beta$-)SiC and various samples of 6H 
($\alpha$-)SiC synthesised for use as diamond replacements.
Though $\beta$-SiC powder and thin film samples have been measured 
in the laboratory previously (see review by Andersen et al. 1999b), this 
study is the first to our knowledge to measure single-crystal (bulk) 
$\beta$-SiC samples in reflectance.
We extracted the real and imaginary parts of the complex refractive index
($n(\lambda) + ik(\lambda)$) from these data using classical
dispersion analyses (Spitzer et al. 1962).
The data presented in 
this work have direct application to carbon-rich AGB stellar
outflows, novae, supernovae, and proto-planetary nebulae.
Given recent Spitzer Space Telescope results for AGB stars and PNe in the 
LMC and SMC (e.g., Speck et al. 2006, Lagadec et al. 2007, 
Zijlstra et al. 2006, Stanghellini et al. 2007), new SiC laboratory 
data are essential to understanding the effect of metallicity on dust 
formation and may also improve our understanding of SiC content of the ISM.


\section{Symmetry analyses of silicon carbide}

Analysing IR data from different variants of SiC requires an
understanding of the relationship between the vibrational modes and structure. 
All varieties of SiC involve tetrahedral linkage of Si and C atoms 
(Taylor \& Jones, 1960).
Two minerals having the same chemical composition but different crystal 
structures are called polymorphs; when structural differences are
due to the stacking of sheets, this is referred to as polytypism.  
About 200 different SiC polytypes exist (Mitra et al. 1969; 
Liu et al. 2004);
it is debated which form most
closely fits the observed feature near 11~$\mu$m in 
C-stars (see review by Speck et al. 1999).
Layers of Si and C atoms can stack into a fcc cubic
crystal (3C, or $\beta$ polytype, SiC; Bechstedt et al. 1997), into hexagonal
layers or into a combination of both cubic and hexagonal shapes
($\alpha$ polytypes).  Schematic diagrams of layering in SiC are
given in the literature (e.g., K\"{a}ckell et al. 1994; Mutschke et al.
1999; Bernatowicz et al. 2006).  The 6H form of $\alpha$-SiC is available 
in large sizes commercially and, thus, is most 
commonly measured in the laboratory. Which polytype of SiC forms 
in space strongly depends on the temperature and gas pressure
within the dust forming region. The 6H form is stable at temperatures up to 
$\sim$~3000~K.  At lower temperatures, 
different polytypes will form, including 4H $\alpha$, 3C $\beta$ and 
2H $\alpha$ (in order of decreasing formation temperature; 
Bernatowicz et al. 2006).  As long as T~$<$~2400~K, $\beta$-SiC 
formation is favoured when condensation takes place in a vacuum.  
$\beta$-SiC will transform into 6H $\alpha$-SiC at
temperatures above $\sim$~2400~K, but it is thermodynamically unlikely
that this process will work in reverse; at best, a few percent
6H $\alpha$-SiC will transform into $\beta$-SiC (I.P. Parkin; private
communication).
Presolar SiC grains contain 3C $\beta$-SiC (80\%), 2H $\alpha$-SiC (3\%), 
and intergrowths of these two forms (17\%) (Daulton et al. 2003).

Structural differences between the polytypes affect their interaction 
with light in two ways.  First, light passing through 
$\beta$-SiC lacks a preferred direction, whereas the layers in $\alpha$-SiC 
cause the interaction of light with Si-C dipoles to differ when
the electric field vector ($\vec{E}$) travels either perpendicular
to the crystal's $\vec{c}$ axis (ordinary ray)
or parallel to $\vec{c}$ (extraordinary ray). 
Second, different numbers of vibrational modes 
are expected for SiC polytypes on the basis of symmetry (Patrick 1968; Feldman et al. 1968).  Because the polytypes are related to one another 
through different stackings of layers, one can infer the number, type, 
and frequency of vibrational modes of any given polytype from the increase 
in size of the crystal's unit cell over that of $\beta$-SiC (i.e., from 
folding the dispersion relations in the first Brillouin zone;
see Burns 1990).
Folding the Brillouin zone of the 3C ($\beta$-)SiC once results in 2H
SiC; folding it three times in succession results in the 6H polymorph.  Phonon
dispersion curves of $\beta$-SiC have been determined using symmetry analysis
and Raman spectroscopic measurements of various polytypes (Nakashima \& 
Harima 1997).
In Raman measurements, the transverse optic (TO) component of 
each IR mode, or minimum in $\epsilon_{2}$, manifests as a peak that
is distinct from the longitudinal optic (LO) component, or maximum 
in Im(1/$\epsilon$).  However, IR activity is complicated by several factors.  
(1) Energy in an IR experiment is absorbed at all frequencies between
the TO and LO components.  (2) Si-C pairs that produce LO activity when 
$\vec{E} \bot \vec{c}$ are the same pairs that produce TO activity when 
$\vec{E} \| \vec{c}$.  (3) Some modes are strong, and some are weak.
Although symmetry considerations do not predict intensities, 
the main Si-C stretch in $\beta$-SiC should dominate the IR spectra of
the various hexagonal and rhombohedral polymorphs. (4) Stacking disorder
may contribute additional modes expected for other polymorphs (e.g., 15R modes
may occur in 6H crystals).  (5) Acoustic overtones and accidental
degeneracies are always possible in IR spectra.
Therefore, Raman frequencies and dispersion curves (Nakashima \&
Harima 1997, Feldman et al. 1968) serve as a useful cross-check for the peak 
wavelengths of our SiC laboratory IR reflectance spectra. 
Our results for 3C- and 6H-SiC peak positions, summarized in 
Table~\ref{table:Raman}, are within 1--3~cm$^{-1}$ of previous determinations 
of the TO and LO modes of SiC (see, e.g., the compilation by 
Mutschke et al. 1999 and references therein).  Comparison to previous 
$n(\nu)$, $k(\nu)$ studies is presented in \S~5.1. 

   \begin{table}
      \caption{IR peak positions of SiC: Raman measurements vs. this study}
         \label{table:Raman}
\begin{minipage}[t]{\columnwidth}
\renewcommand{\footnoterule}{}  
\begin{tabular}{lccc}     
\hline\hline
            Peak & Nakashima \& Harima (1997)$^{\mathrm{a}}$ & \multicolumn{2}{ |c }{this work} \\
            Type & Raman $\lambda$ ($\mu$m) & 3C IR $\lambda$ & 6H IR $\lambda$ \\
 \hline\hline
            LO & 10.28 & 10.28 & $\bot$ 10.26   \\
            LO & 10.36 & --- & $\|$ 10.23 \\
               &       &     & $\bot$ 10.31 \\
            folded & 10.64 & --- & --- \\
            folded & 11.25 & --- & 11.3 \\
            LA $+$ TA & --- & 11.35 & $\|$ 11.32 \\
            folded & 11.96 & --- & $\|$ 11.96 \\
            TO & 12.54 & 12.54 & $\bot$ 12.54 \\
            folded & 12.67 & --- & $\|$ 12.69\\
            folded & 12.95 & --- & --- \\
            folded & 13.04 & --- & --- \\
            LA & 16.26 & --- & --- \\
            TA & 37.59 & --- & --- \\
            \hline
 \hline	    
 \end{tabular}
 \end{minipage}
\begin{list}{}{}
\item[$^{\mathrm{a}}$] 
Additional Raman study references given in Mutschke et al. (1999).
\end{list}
   \end{table}

\section{Experimental methods}

\subsection{Sample characteristics}

Grain properties and manufacturers' information for the SiC
samples studied here are presented in Table~\ref{table:1}.  
We verified sample polytype by optical 
microscopy and by the spectroscopic results below.  Some samples 
were highly pure (Table~\ref{table:1}).  
Though we cannot rule out small departures from non-stoichiometry (e.g., 
Kimura et al. 2006), impurities at the few per cent 
level would be required to affect band positions in the mid- to far-IR.
The colours of the $\alpha$-SiC samples are associated 
with impurities which should not affect the main spectral 
band. Under magnification of x50, the samples are homogeneous in 
colour and, thus, in estimated impurities.
The samples lack inclusions but present crystal growth sectors 
(i.e., planes parallel to $\vec{c}$). 

Hexagonal polymorphs of SiC grow as plates perpendicular to the crystal plane
(001).  This orientation was confirmed using sample morphology and 
optical microscopy.  For $\alpha$-SiC, we created $\vec{a}$-$\vec{a}$ 
plates and $\vec{a}$-$\vec{c}$ plates by grinding and polishing parallel 
or perpendicular to the large faces of our samples, respectively.  
Because SiC is an extremely hard mineral (9.1--9.5 on the Mohs scale at 
$20^{\circ}$~C), with layering and growth sectors, mirror surfaces were 
difficult to attain.  Thus, the measured reflectance will 
be lower than the true, absolute reflectance.   

\subsection{Laboratory IR spectroscopic measurements}

Room temperature (18--19$^{\circ}$C) IR specular reflectance spectra were 
acquired at near-normal incidence (i.e., the beam passes through the 
microscope at an angle of $=$~90$\pm$10$^{\circ}$) using a Spectra-Tech 
Fourier transform infrared (FT-IR) spectrometer\footnote{Spectra-Tech Inc. 
(Thermo Electron Corp.), Stamford, CT, USA.} microscope in an evacuated 
Bomem DA 3.02 Fourier transform spectrometer\footnote{Bomem Inc., 
Quebec, Canada.}. Resolutions of 1~cm$^{-1}$ 
 (mostly for the mid-IR) or 2~cm$^{-1}$ resolution 
 (far-IR) suffice to separate peaks for solid samples at room temperature.  
 Instrumental accuracy is $\sim$~0.01~cm$^{-1}$.  For the 
 $\vec{E} \| \vec{c}$ polarization of 6H only, we used a specular 
 reflection device instead of the microscope, wherein the incident beam 
 strikes the centre of the sample at 30$^{\circ}$ to the normal.  We 
 used the ``S'' polarization, where the direction of the electric field 
 is parallel to the line defined by the two mirror planes.  The beam 
 size was 600~$\mu$m in the microscope and $\sim$~1~mm in the specular 
 reflectance device.  A Si-bolometer and a coated mylar beamsplitter were used 
 for the far-IR, $\sim$~50 to 650~cm$^{-1}$ ($\sim$~200--15.0~$\mu$m).  A 
 KBr beamsplitter and a liquid-nitrogen-cooled HgCdTe detector were 
 used for the mid-IR, 450--4000~cm$^{-1}$ ($\sim$~22--2.5~$\mu$m).  A 
 gold mirror (100\% reflection assumed; 98\% average reflection measured) was 
used as the reference.  2000 scans yielded a reasonable noise level. 

The laboratory reflectance spectra 
in Fig.~\ref{allEparc} and Fig.~\ref{allEperpc} have been merged, 
corrected for artifacts and rescaled from the raw 
spectra.  Where reflectance spectra were available for both wavelength 
regions, mid- and far-IR reflectance intensities were scaled to 
match in the region of overlap and merged.  Because back reflections 
(e.g., from growth sectors, Hofmeister et al. 2003) increase 
apparent reflectance, the segment with the lowest reflectance 
above 1000~cm$^{-1}$ was presumed to be correct. 
For the $\vec{a}$-$\vec{a}$ sections, we collected spectra from 
several areas and present 
the spectrum with the highest reflectance at the peak center 
(i.e., best polished surface).  For our $\alpha$-SiC,
the maximum absolute reflectance measured is low ($R$~$=$~0.72) 
due to surface imperfections.  We estimate the effect of errors 
in $R$ on our data analysis and compare peak shapes 
by scaling all reflectance spectra to 92--99\% maximum reflectance 
(cf. Spitzer et al. 1959b;  
Il'in et al. 1972; Zorba et al. 1996; Goncharenko et al. 1996).
Our (blue-gray) $\alpha$-SiC hopper crystal sample was large enough 
to provide good data from the $\vec{a}$-$\vec{c}$ section, but its 
structure of intergrown, stacked crystals caused artifacts in the 
reflectance spectrum.  The $\vec{a}$-$\vec{a}$ plates we studied were 
too thin to allow us to collect data from the $\vec{a}$-$\vec{c}$ edges.  
The size of our mossanite sample was marginally large enough for 
a $\vec{a}$-$\vec{c}$ sample but is comparable in size to the beam 
diameter and thus our apparatus may not have sampled the same areas 
from the reference mirror and the moissanite sample.

\subsection{IR Data Analysis}

To avoid presenting results that are affected back reflections, we compared
 reflectance $R_{\nu}$ in the limit that $k$~$\rightarrow$~0 at 
 visible wavelengths,

\begin{equation}
R_{\nu} \approx \frac{(n_{\nu} - 1)^{2}}{(n_{\nu} + 1)^{2}}.
\label{rnueq}
\end{equation}

\noindent
with literature values of the index of refraction determined through 
microscopy in white or yellow light (either $n(\lambda$~$=$~467 nm) 
or $n$($\nu$~$=$~22\,000~cm$^{-1}$) 
($\lambda$~$=$~467 nm or $\nu$~$=$~22\,000~cm$^{-1}$, respectively,
Table~\ref{table:1}). If the reflectance at $\nu$~$=$~4000~cm$^{-1}$ 
was less than $R_{white}$ by a few percent, then back reflections 
did not affect the data.

We used Kramers-Kronig analyses (Fahrenfort 1961; Roessler 1965) to determine 
starting estimates of peak positions and widths 
and followed up with classical dispersion analyses to provide 
robust $n(\nu)$ and $k(\nu)$ values.  $\beta$-SiC has poorly 
resolved peaks;
this is problematic for a Kramers-Kronig calculation, but 
not for classical dispersion analysis (e.g., Giesting \& Hofmeister 2002).
 Values of $n$ and $k$ from our best fits are presented in 
 Figs.~\ref{alphaEparc1osc}b-\ref{betawafer98}b and online at the CDS 
(Tables~3--7).

\begin{table*}
\caption{Experimental samples: Manufacturer information and references}
\label{table:1}
\begin{minipage}[t]{2\columnwidth}
\renewcommand{\footnoterule}{}  
\begin{tabular}{lcccccc}     
\hline\hline
Polytype & Mineral Name & Grain Size
& Manufacturer & Refractive Index & Comments \\
 \hline\hline
 $\alpha$-SiC & synthetic & diam. = 6.5~mm
           & Charles \& & n$_{o}$~$=$~2.654\footnote{Indices of refraction for
	   the O-ray ($n_{o}$) and E-ray ($n_{e}$) given for
	   white light (Gaines et al. 1997) or IR $\lambda$ (Goldberg et al. 2001).}
          & round brilliant \\
 & moissanite & & Colvard, Ltd$^{TM}$ & n$_{e}$~$=$~2.967 & cut gem \\
 $\alpha$-SiC & 6H-SiC & 2~$\mu$m powder; & Alfa/Aesar & n$_{o}$~$=$~2.55, n$_{e}$~$=$~2.59 (IR)
    & hexagonal plates \\
         & (gray, amber) & surf. area $=$~9--11~m$^{2}$~g$^{-1}$
            & (Lot \# C19H06) & n$_{o}$~$=$~2.5531$+$(3.34$\cdot$10$^{4}$)$\cdot \lambda^{-2}$, & purity: 99.8\% \\
              & & & & n$_{e}$~$=$~2.5852$+$(3.68$\cdot$10$^{4}$)$\cdot 
	\lambda^{-2}$ \footnote{Values for visible light (467~nm~$<$~
	$\lambda$~$<$~691~nm at $T$~$=$~300~K; Schaffer \& Naum 1969, 
	Schaffer 1971).
	Formulae assume $\lambda$ in nanometers.} & metals basis \\
   $\alpha$-SiC & 6H-SiC (blue-gray); & several mm & unknown & n$_{o}$~$=$~2.654$^{\mathrm{a}}$ & Hopper crystal; \\
         & synthetic & per crystal & & n$_{e}$~$=$~2.967 & (i.e., intergrown \\
	 & carborundum & & & & crystals) \\
   $\alpha$-SiC & 6H-SiC (green, & 4--8~mm wide, & Morion Co. & n$_{o}$~$=$~2.654$^{\mathrm{a}}$ & green: layered, hexagonal, flat \\
         & yellow) & 0.25--1.20~Car. each & & n$_{e}$~$=$~2.967 & yellow: single crystal\\
   $\beta$-SiC & 3C-SiC wafer & diam$=$~5~$\mu$m & Rohm \& Haas,
        & n$_{o, e}$~$=$~2.55$^{\mathrm{b}}$ & CVD wafer \\
        & (fcc cubic) & & Advanced Materials & n$_{o, e}$~$=$~2.55378$+$(3.417$\cdot$10$^{4}$)$\cdot \lambda^{-2}$ $^{\mathrm{c}}$ & purity: $\geq$~99.9995\% \\
        & & & (Grade SC-001) & \\
   $\beta$-SiC & 3C-SiC gray & diam. $\sim$~2.5--25~$\mu$m  & Superior Graphite 
       & n$_{o, e}$~$=$~2.55$^{\mathrm{b}}$ & equant (spherical) \\
        & (fcc cubic) & & & n$_{o, e}$~$=$~2.55378$+$(3.417$\cdot$10$^{4}$)$\cdot \lambda^{-2}$ $^{\mathrm{c}}$ & chips \\

 \hline
 \hline
 $$
 \end{tabular}
 \end{minipage}
 \end{table*}

\subsubsection{Classical dispersion analysis}

Reflection and transmission by bulk media can be described 
using the complex refractive index ($n - ik$) which in turn are related to 
the complex dielectric function ($\epsilon_{1} + i\epsilon_{2}$).

\begin{equation}
n = ( \frac{1}{2} ( ( \epsilon_{1}^{2} + \epsilon_{2}^{2} )^{-1/2} + \epsilon_{1} ) )^{-1/2}
\label{angeq3}
\end{equation}

\begin{equation}
k = ( \frac{1}{2} ( ( \epsilon_{1}^{2} + \epsilon_{2}^{2} )^{-1/2} - \epsilon_{1} ) )^{-1/2}
\label{angeq4}
\end{equation}

\begin{equation}
\epsilon_{1} = n^{2} - k^{2}
\label{angeq1}
\end{equation}

\begin{equation}
\epsilon_{2} = 2nk
\label{angeq2}
\end{equation}

These quantities may be obtained by fitting the laboratory reflectance spectra 
using classical dispersion analysis (e.g., Spitzer et al. 1962).  This 
method treats the vibrations produced by the interaction of light with 
the sample as damped harmonic oscillators (i.e., the peaks in 
$\epsilon_{2}$ and Im(1/$\epsilon$) have Lorentzian shapes, cf. Wooten 1972). 

We constructed synthetic reflectance spectra from three parameters: (1) 
the TO peak positions ($\nu_{i}$) determined from the maxima 
in $\epsilon_{2}(\nu)$, the full width at half-maximum (FWHM$_{i}$) of each 
peak in $\epsilon_{2}(\nu)$, and the oscillator strength 
$f_{i}$~$=$2 FWHM$_{i} \sigma_{max} / \nu_{i}^{2}$, where the 
conductivity $\sigma(\nu) = \nu \epsilon_{2}(\nu)/2$.  The light 
angle of incidence $\Phi$ is accounted for after Jackson (1975):

\begin{equation}
r_{s\Phi} = \sqrt{R_{meas}} = \frac{ cos\Phi - \sqrt{n^{2} - k^{2} + 2ink - sin^{2}\Phi} }{ cos\Phi + \sqrt{n^{2} - k^{2} + 2ink - sin^{2}\Phi} }.
\end{equation}

The equations for the components of the dielectric function for $m$ oscillators 
using 
the damping coefficient $\Gamma_{i} = 2\pi$FWHM$_{i}$ are

\begin{equation}
\epsilon_{1} = \epsilon_{\infty} + \sum_{j=1}^{m} \frac{ 4\pi^{2}f_{j} (\nu_{j}^{2} - \nu^{2} ) }{ 4\pi^{2}( \nu_{j}^{2} - \nu^{2})^{2} + \Gamma_{j}^{2}\nu^{2} },
\label{epsilon1eq}
\end{equation}

\noindent
where $\epsilon_{\infty}$ is $n^{2}$ in the visible, and 

\begin{equation}
\epsilon_{2} = \sum_{j=1}^{m} \frac{ 2 \pi f_{j} \nu_{j}^{2} \Gamma_{j} \nu }{ 4\pi^{2}(\nu_{j}^{2} - \nu^{2})^{2} + \Gamma_{j}^{2}\nu^{2} }.
\label{epsilon2eq}
\end{equation}

\noindent
The absorption coefficient $A$ is calculated from 

\begin{equation}
A(\nu) = \frac{2\pi\nu\epsilon_{2}(\nu)}{n(\nu)} = 4 \pi \nu k(\nu).
\label{abscoeffeq}
\end{equation}

We note that Mutschke et al. (1999) have calculated spectral absorption and 
emission cross-sections for different particle shapes using ad hoc damping 
coefficients $\gamma$ for 6H-SiC.  Mutschke et al. (1999) provide a 
simplified version of Eq.~7 that is appropriate for spectra arising from 
one oscillator (their eq. 1) and further relate the strength of a single 
oscillator to $\nu_{LO}^{2} - \nu_{TO}^{2}$ (their eq. 2).  Because 
multiple oscillators occur for certain polarizations of SiC and are 
seen in all of our samples, we instead use Eqs.~7--9 to constrain peak 
widths and oscillator strengths.

 \begin{figure}
 \centering
    \includegraphics[width=8cm]{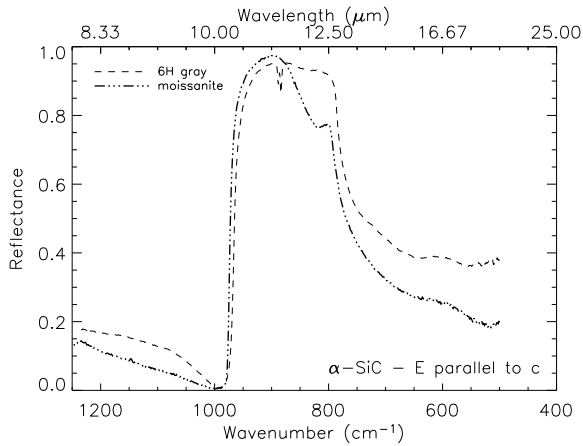}
    \caption{Mid- and mid$+$far-IR laboratory specular reflectance as a 
    function of wavenumber (wavelength) for $\vec{E} \| \vec{c}$ faces 
    of $\alpha$-SiC: moissanite (dash-dot line) and gray 6H 
    (dashed line).  Laboratory values scaled to 95\% maximum reflectance from 
    35\% (moissanite) and 45\% (gray 6H).}  
 \label{allEparc}
 \end{figure}

 \begin{figure}
 \centering
    \includegraphics[width=8cm]{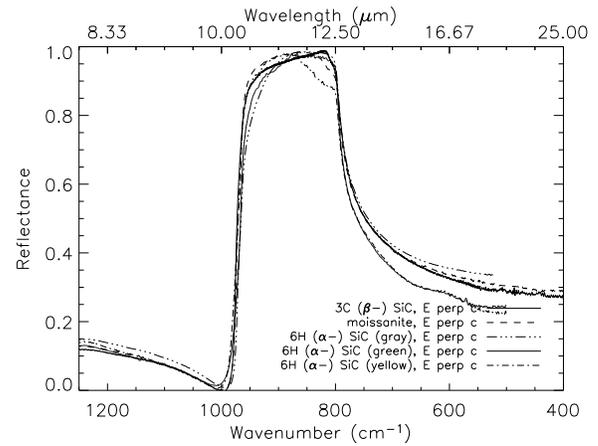}
  \caption{Mid- and mid$+$far-IR laboratory specular reflectance 
    for $\vec{E} \bot \vec{c}$ faces of $\beta$-SiC 
    (dashed line), and two $\alpha$-SiC samples: 
    moissanite (solid line) and gray 6H 
   (dash-dot line).  A mode occurs at $\sim$~965~cm$^{-1}$;
   polish effects account for the differences in shape of the main peak.  
   Values scaled from 72\% ($\beta$-SiC), 95\% (moissanite), 84\% (gray 6H), 
   76\% (green 6H) and 84\% (yellow 6H) to 98\% maximum reflectance.}
\label{allEperpc}
\end{figure}

\section{Results}

\subsection{Laboratory reflectance spectra}

Laboratory reflectance spectra of $\alpha$- and $\beta$-SiC are 
presented in Figs.~\ref{allEparc}--\ref{allEperpc}; peak positions are 
given in cm$^{-1}$ wavenumber, or 10$^{4}$/($\lambda$ in $\mu$m).  
Reflectance spectra of $\beta$-SiC show
a large, broad feature extending from the TO
position near 797.5~cm$^{-1}$ to a LO mode near 973~cm$^{-1}$, 
consistent with Raman frequencies measured by Feldman et al. (1968). 
Because the sum of transverse and longitudinal acoustic modes
at $\nu_{TA} + \nu_{LA} =$~876~cm$^{-1}$ falls between the LO and
TO modes of $\beta$-SiC, a resonance is possible near that position.  
For $\alpha$-SiC, from zone folding and Raman measurements of the dispersion, 
IR modes for $\vec{E} \| \vec{c}$ could also occur at 965, 940, 889, 
and 836~cm$^{-1}$ and in $\vec{E} \bot \vec{c}$ at 797, 789, 772, and 
767~cm$^{-1}$.  As in $\beta$-SiC, the strong TO mode in $\alpha$-SiC 
occurs near 797~cm$^{-1}$ in $\vec{E} \bot \vec{c}$ and at 
$\sim$ 789~cm$^{-1}$ for $\vec{E} \| \vec{c}$ in $\alpha$-SiC.  
Modes at lower frequencies are not apparent, i.e., these are too weak 
for resolution in reflectivity data, if they exist.

We fitted the laboratory reflectance spectra via classical dispersion analysis 
to characterize the position, FWHM, and oscillator strength 
parameters of the main peak and also to resolve the structure near the TO 
and LO positions in both SiC polytypes.  This structure does not occur 
near the expected peak positions and appears to be due to physical 
optics effects, as discussed in 
\S~4.2.  The suspect features may be described as a divot in the reflectance spectrum near the TO mode for both polarizations of the $\alpha$-SiC (moissanite and 6H gray) samples, in $\vec{E} \bot \vec{c}$ polarization only for the yellow $\alpha$-SiC sample, and also in the $\beta$-SiC wafer sample.  In $\vec{E} \bot \vec{c}$ for the 6H gray and, to some extent, the green $\alpha$-SiC reflectance spectra, the slope near the LO mode is more shallow than that of the yellow $\alpha$-SiC and moissanite reflectance spectra and may also be a spectral artifact.

\subsection{Optical constants: $n(\nu)$, $k(\nu)$}

\begin{figure}
  \centering
    \begin{minipage}[b]{0.5\textwidth}
       \includegraphics[width=\textwidth]{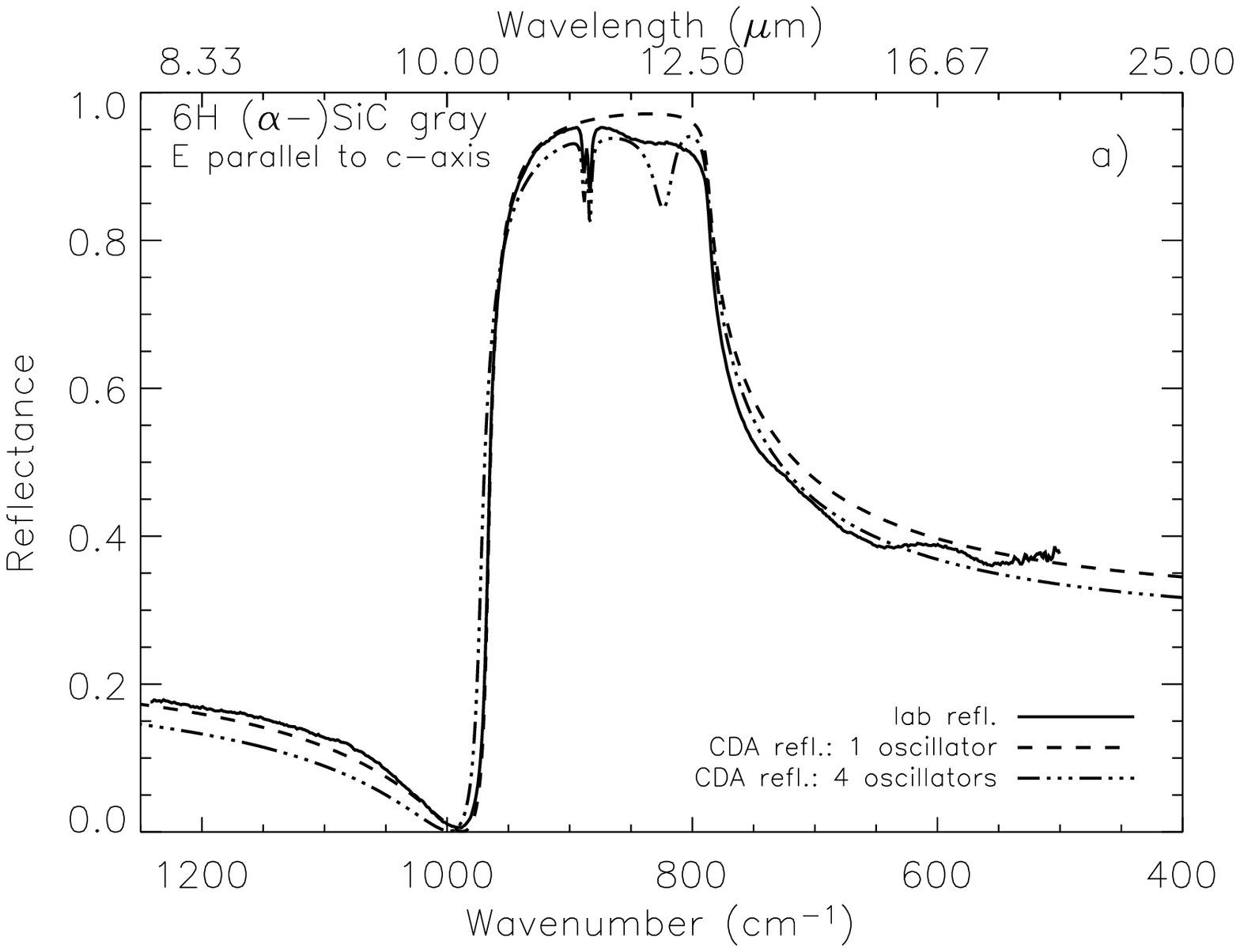}
    \end{minipage}\\
    \begin{minipage}[b]{0.5\textwidth}
       \includegraphics[width=\textwidth]{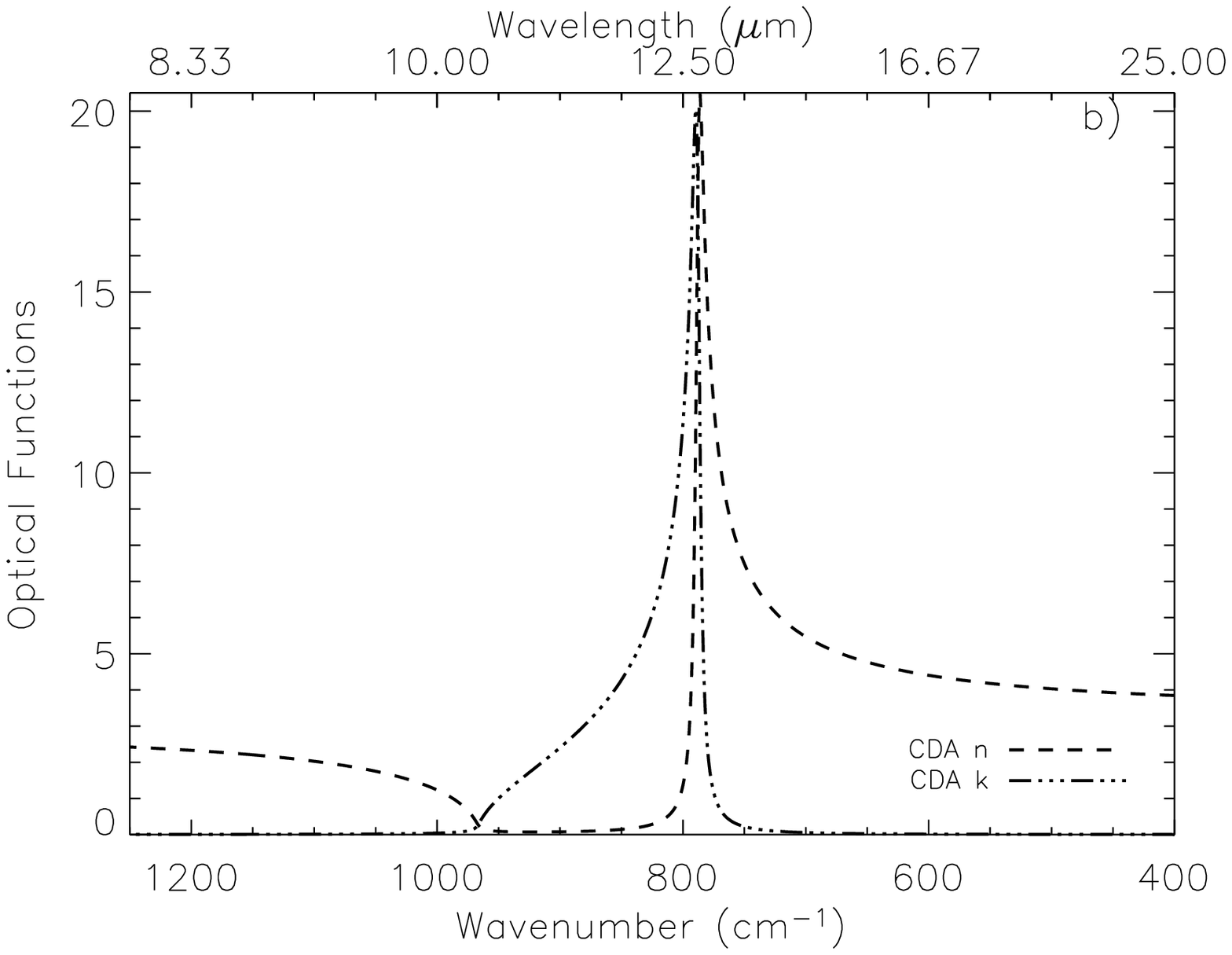}
    \end{minipage}
    \caption{Reflectivity at near-normal incidence
    of 6H gray $\alpha$-SiC 
    ($\vec{E} \| \vec{c}$ 
    orientation) and derived functions from
  classical dispersion analysis. 
  (a) Laboratory reflectivity spectrum (solid line, scaled to 
  95\% maximum reflectance), calculated reflectivity fitted by 1 oscillator
    ($\nu_{1}$~$=$~787.5~cm$^{-1}$, FWHM$_{1}$~$=$~5.50~cm$^{-1}$, oscillator
       strength $f_{1}$~$=$~4.45)
  (long dashed line), and fitted by 4 oscillators
  ($\nu_{1}$~$=$787.5~cm$^{-1}$, FWHM$_{1}$~$=$~6.60~cm$^{-1}$, 
  $f_{1}$~$=$~3.57; $\nu_{2}$~$=$~825.0~cm$^{-1}$, 
   FWHM$_{2}$~$=$~20.0~cm$^{-1}$, $f_{2}$~$=$~0.3; 
   $\nu_{3}$~$=$~883.7~cm$^{-1}$, 
   FWHM$_{3}$~$=$~2.5~cm$^{-1}$, $f_{3}$~$=$~0.00475; 
   $\nu_{4}$~$=$~888.5~cm$^{-1}$, 
   FWHM$_{4}$~$=$~3.3~cm$^{-1}$, $f_{4}$~$=$~0.00425) (short dashed line).
  (b) Real and imaginary parts of
  the complex index of refraction for the 1 oscillator fit: 
  $n$ (solid line) and $k$ (dashed line).}
  \label{alphaEparc1osc}
 \end{figure}

\begin{figure}
  \centering
   \begin{minipage}[b]{0.5\textwidth}
       \includegraphics[width=\textwidth]{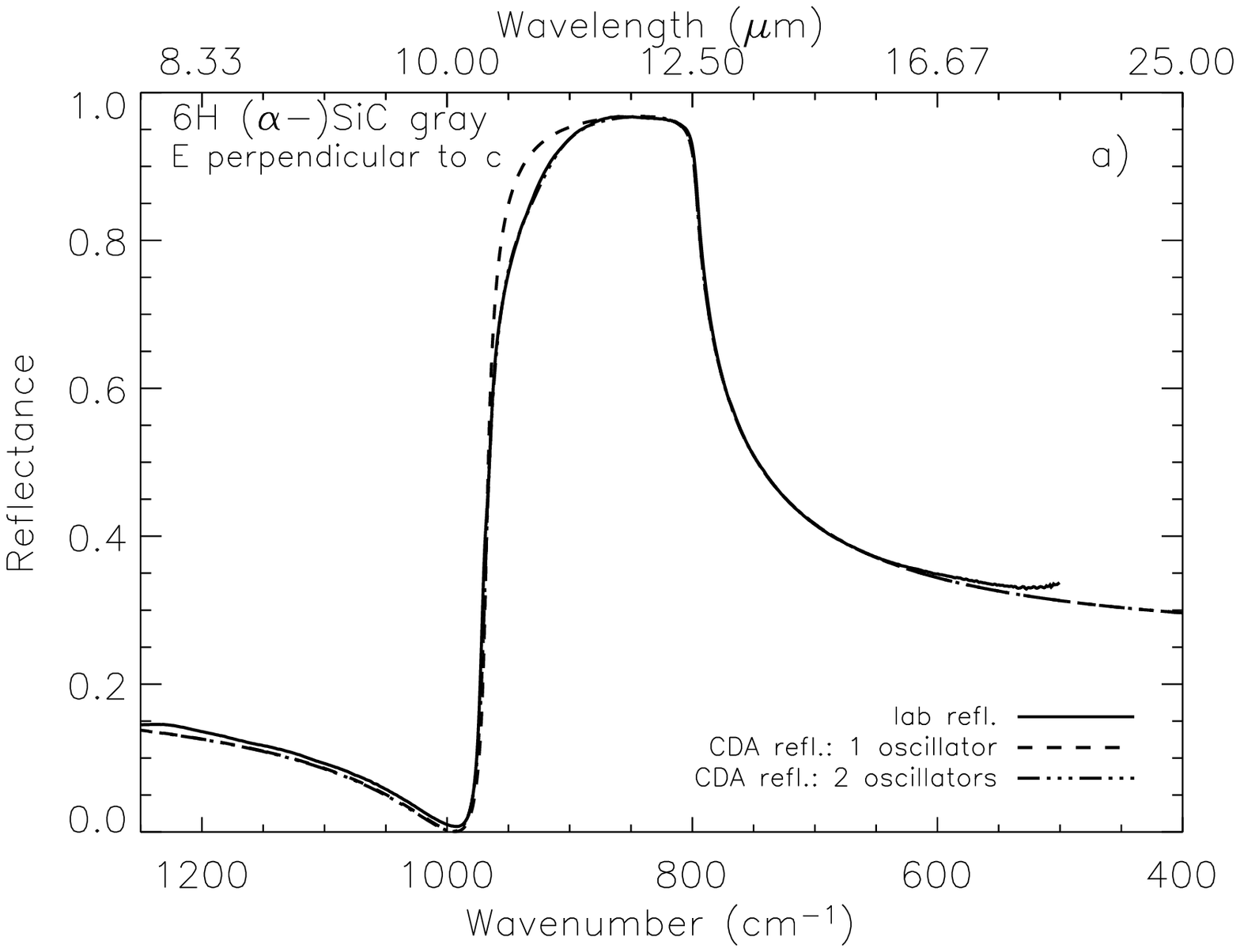}
   \end{minipage}\\
  \begin{minipage}[b]{0.5\textwidth}
       \includegraphics[width=\textwidth]{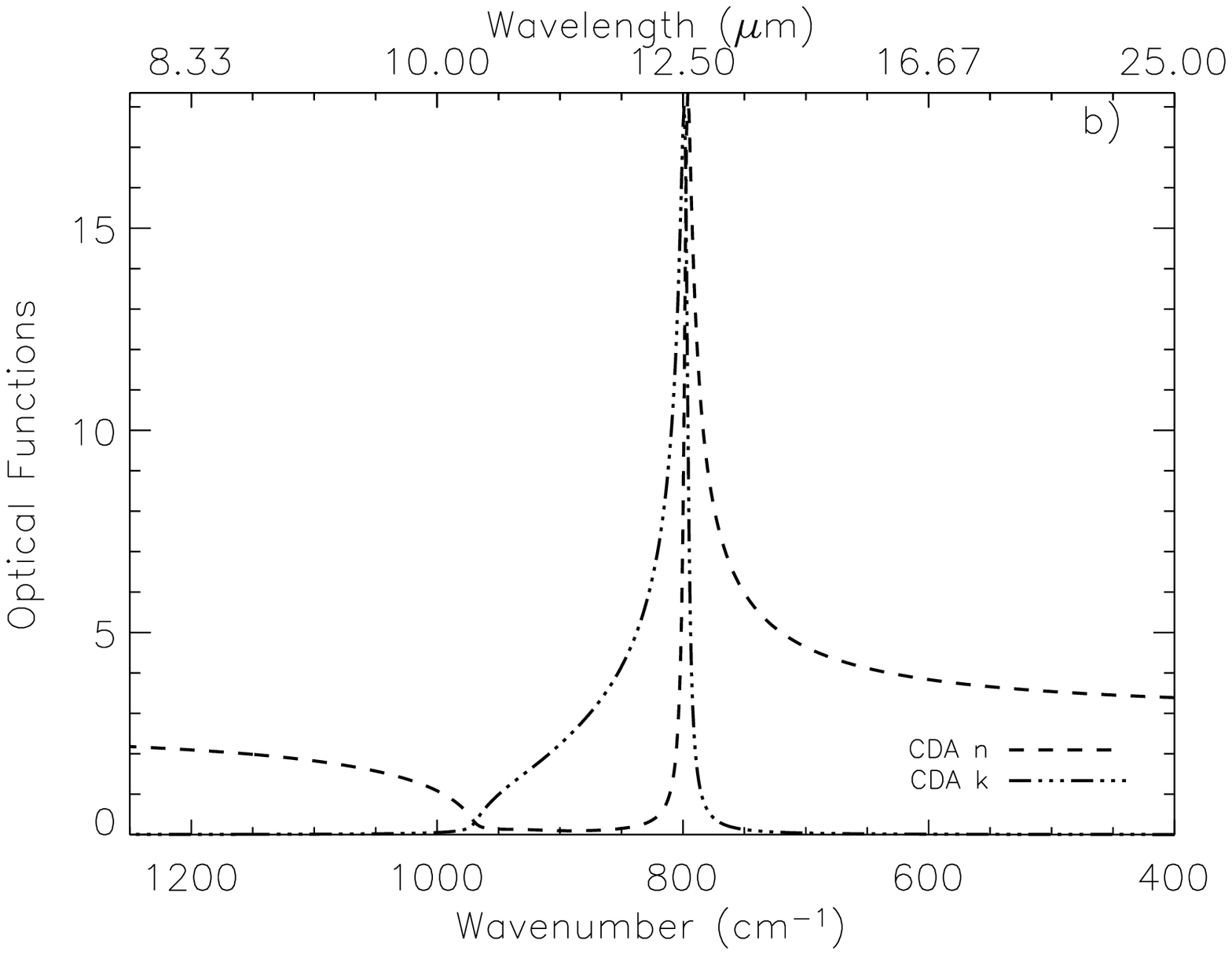}  
  \end{minipage}
  \caption{Reflectivity at near-normal incidence
   of 6H gray $\alpha$-SiC ($\vec{E} \bot \vec{c}$
   orientation) and derived $n$ and $k$.
  (a) Laboratory reflectivity spectrum (solid line,
   scaled to 97\% maximum reflectance),
   calculated reflectivity fitted by 1 oscillator
   ($\nu$~$=$~797.5~cm$^{-1}$, FWHM~$=$~5.30~cm$^{-1}$, 
   $f$~$=$~3.33) (long dashed line), and by 2 oscillators
   ($\nu_{1}$~$=$~797.5~cm$^{-1}$, FWHM$_{1}$~$=$~5.20~cm$^{-1}$, 
   $f_{1}$~$=$~3.32; $\nu_{2}$~$=$~928.0~cm$^{-1}$, 
   FWHM$_{2}$~$=$~65.0~cm$^{-1}$, $f_{2}$~$=$~0.013) (short dashed line).
 (b) $n$ (solid line) and $k$ (dashed line) for the 2 oscillator fit.}
 \label{alphaEperpc1osc}
 \end{figure}

\begin{figure}
  \centering
   \begin{minipage}[b]{0.5\textwidth}
    \includegraphics[width=\textwidth]{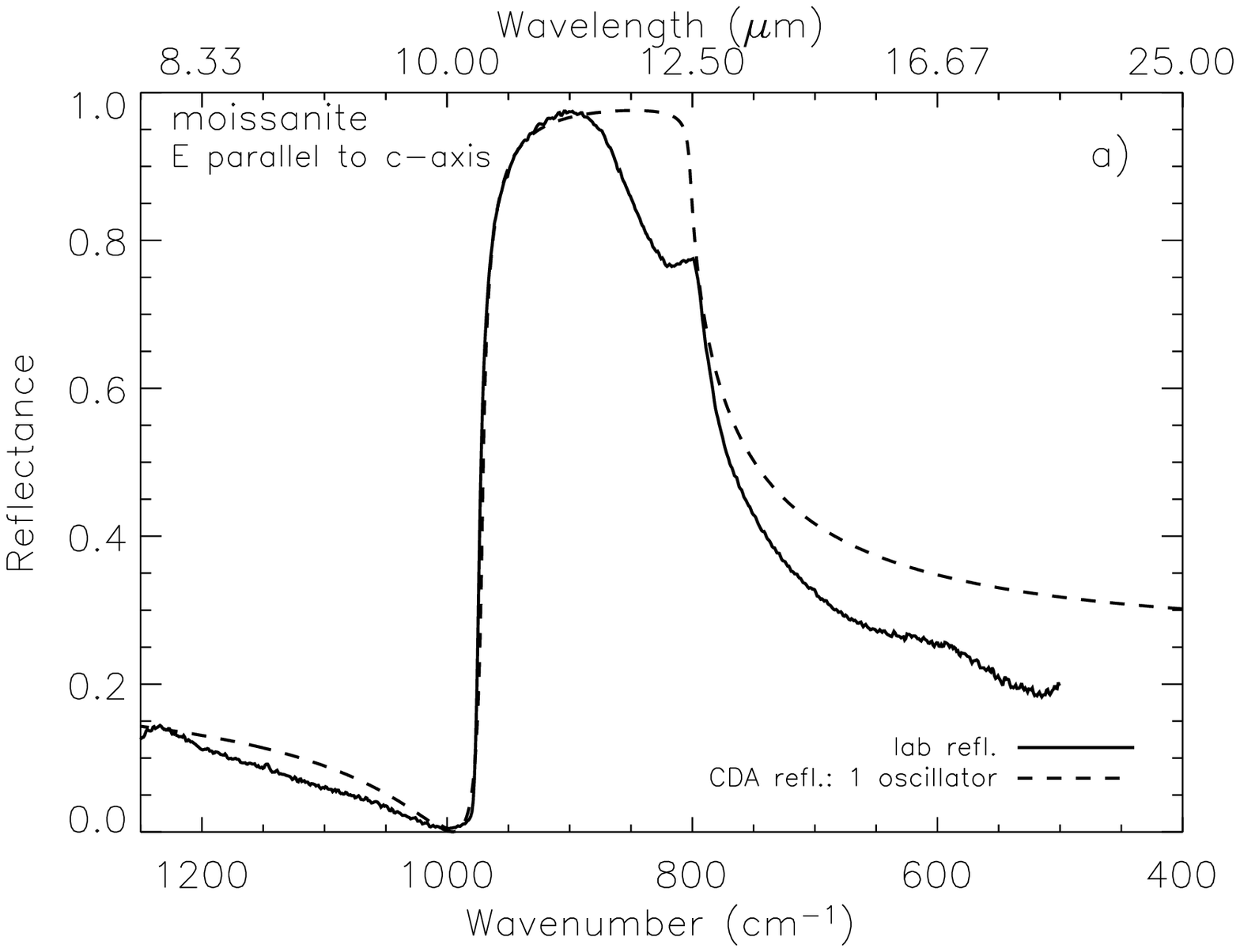}
  \end{minipage}\\
   \begin{minipage}[b]{0.5\textwidth}
     \includegraphics[width=\textwidth]{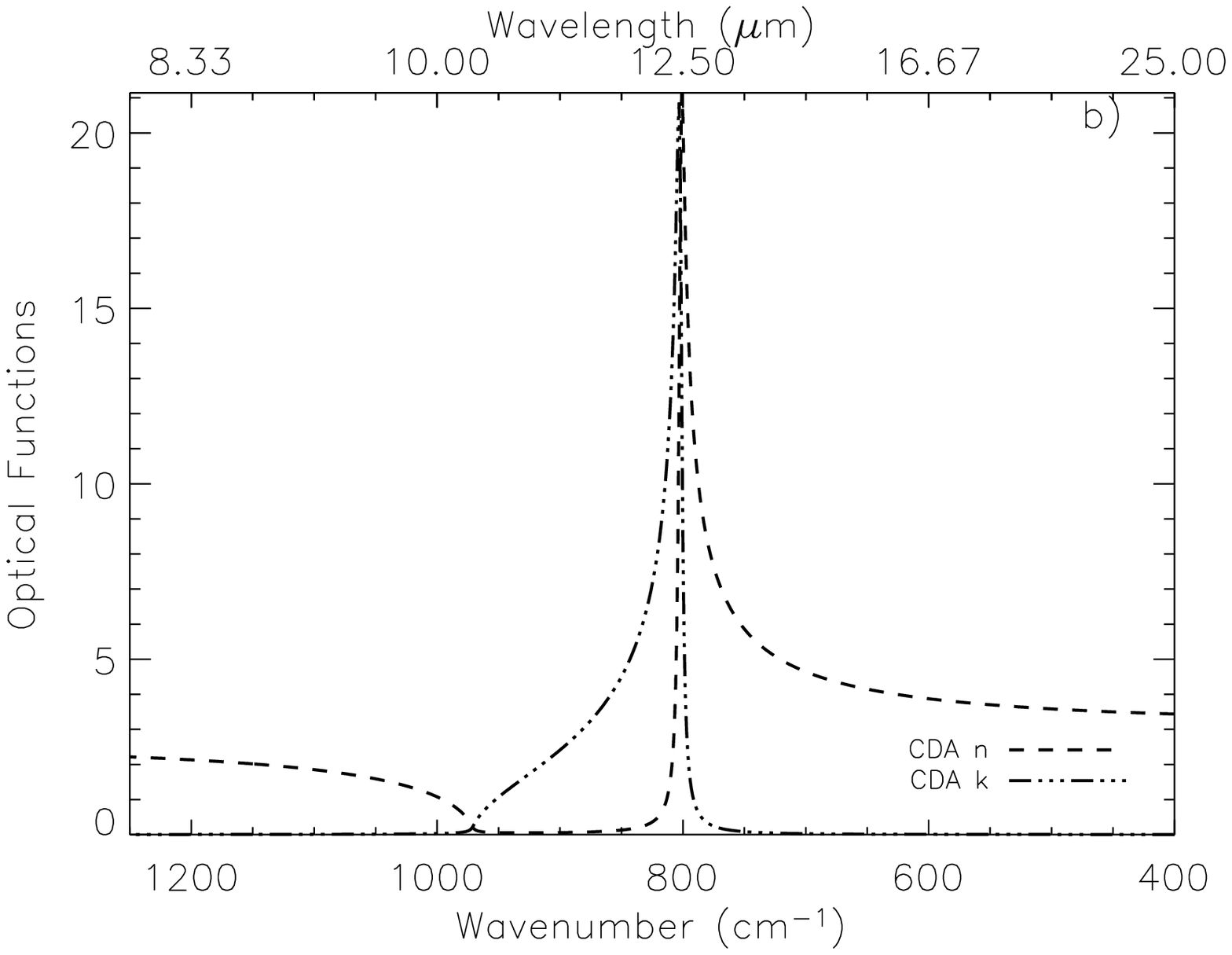}
   \end{minipage}
   \caption{Reflectivity at near-normal incidence of 
   moissanite ($\alpha$-SiC, $\vec{E} \| \vec{c}$ orientation) and derived 
   $n$ and $k$ fitted by 1 oscillator
   ($\nu$~$=$~802.0~cm$^{-1}$, FWHM~$=$~4.0~cm$^{-1}$, 
   $f$~$=$~3.4).  (a) Laboratory reflectivity spectrum (solid line,
    scaled to 97\% maximum reflectance)
    and calculated reflectivity (dashed line).
  (b) $n$ (solid line) and $k$ (dashed line).}
  \label{moissEparc1osc}
  \end{figure}

\begin{figure}
  \centering
   \begin{minipage}[b]{0.5\textwidth}
       \includegraphics[width=\textwidth]{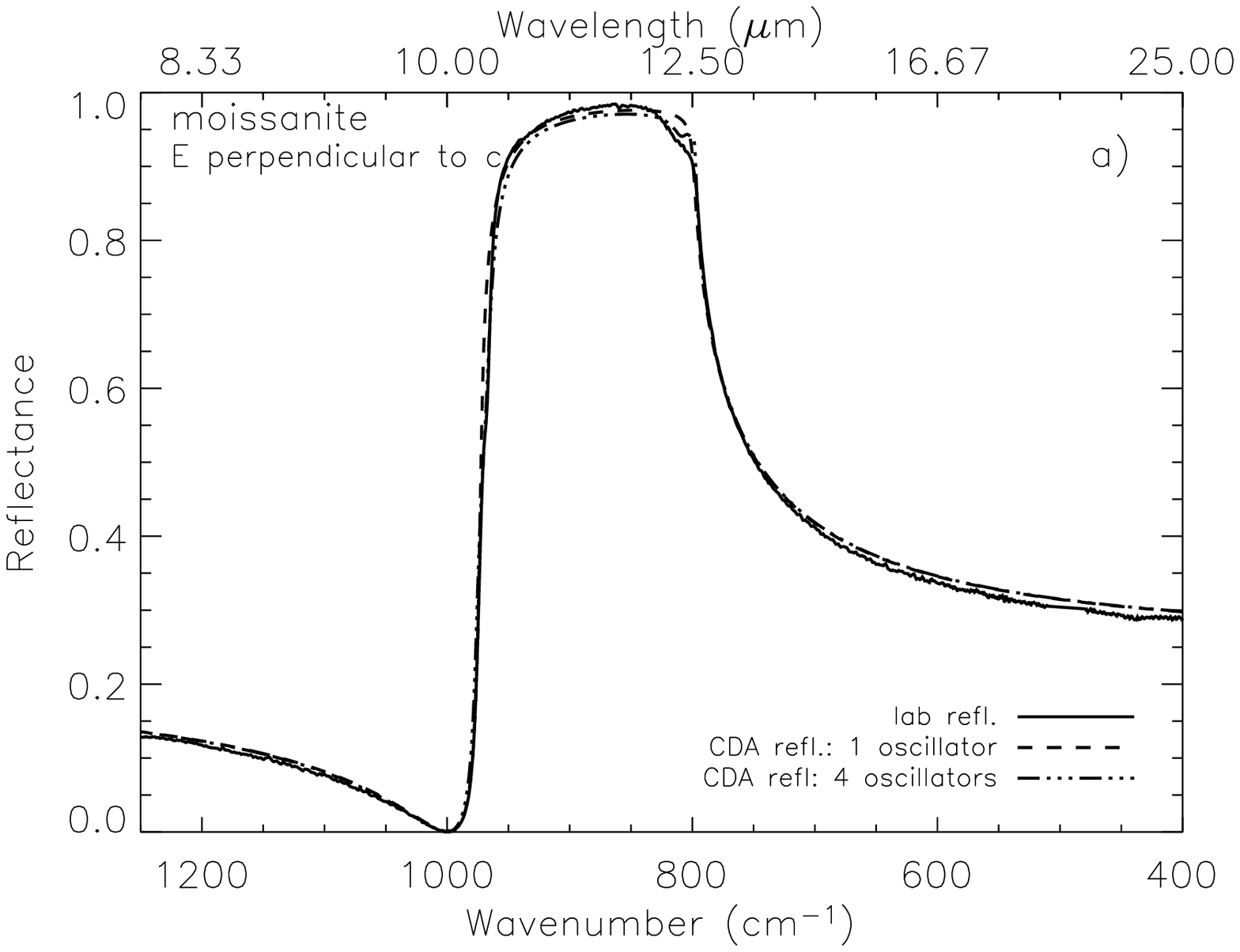}
   \end{minipage}\\
 \begin{minipage}[b]{0.5\textwidth}
       \includegraphics[width=\textwidth]{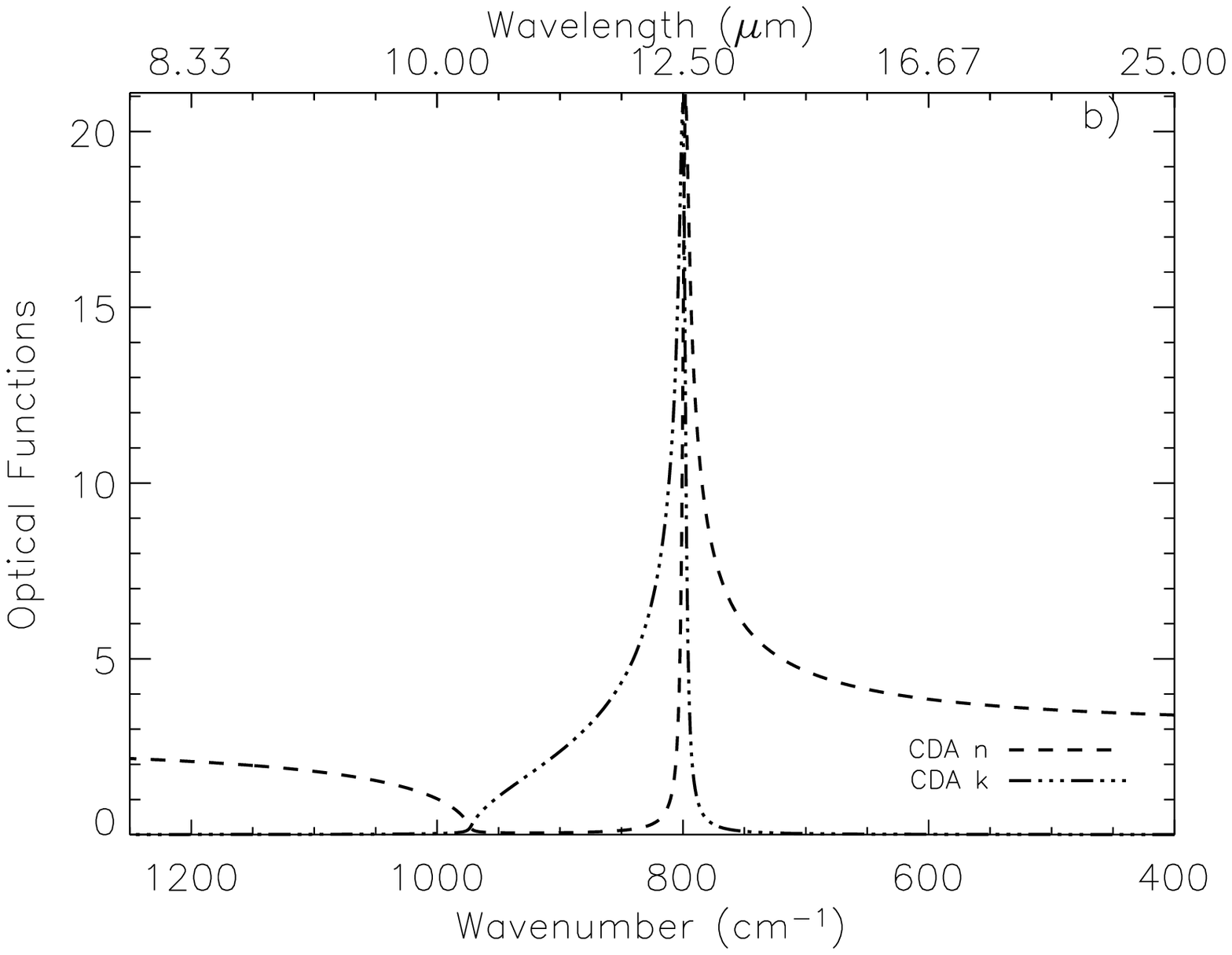}
  \end{minipage}
  \caption{Reflectivity at near-normal incidence of moissanite 
  ($\alpha$-SiC, $\vec{E} \bot \vec{c}$ orientation) and derived 
  $n$ and $k$.  
  (a) Laboratory reflectivity spectrum (solid line, 
   scaled to 97\% maximum reflectance), calculated reflectivity
  fitted by 1 oscillator
  ($\nu$~$=$~799.0~cm$^{-1}$, FWHM~$=$~4.0~cm$^{-1}$, 
  $f$~$=$~3.4) (long dashed line), and by 4 oscillators
  ($\nu_{1}$~$=$~797.5~cm$^{-1}$, FWHM$_{1}$~$=$~1.75~cm$^{-1}$, 
  $f_{1}$~$=$~2.04; $\nu_{2}$~$=$~799.0~cm$^{-1}$, 
  FWHM$_{2}$~$=$~7.0~cm$^{-1}$, 
   $f_{2}$~$=$~1.10; $\nu_{3}$~$=$~808.0~cm$^{-1}$, 
   FWHM$_{3}$~$=$~14.0~cm$^{-1}$, 
   $f_{3}$~$=$~0.28; $\nu_{4}$~$=$~970.0~cm$^{-1}$, 
   FWHM$_{4}$~$=$~12.5~cm$^{-1}$, $f_{4}$~$=$~0.0010) (short dashed line).
   (b) $n$ (solid line) and $k$ (dashed line) for the 1 oscillator fit.}
  \label{moissEperp1osc}
  \end{figure}

\begin{figure}
    \centering
    \begin{minipage}[b]{0.5\textwidth}
       \includegraphics[width=\textwidth]{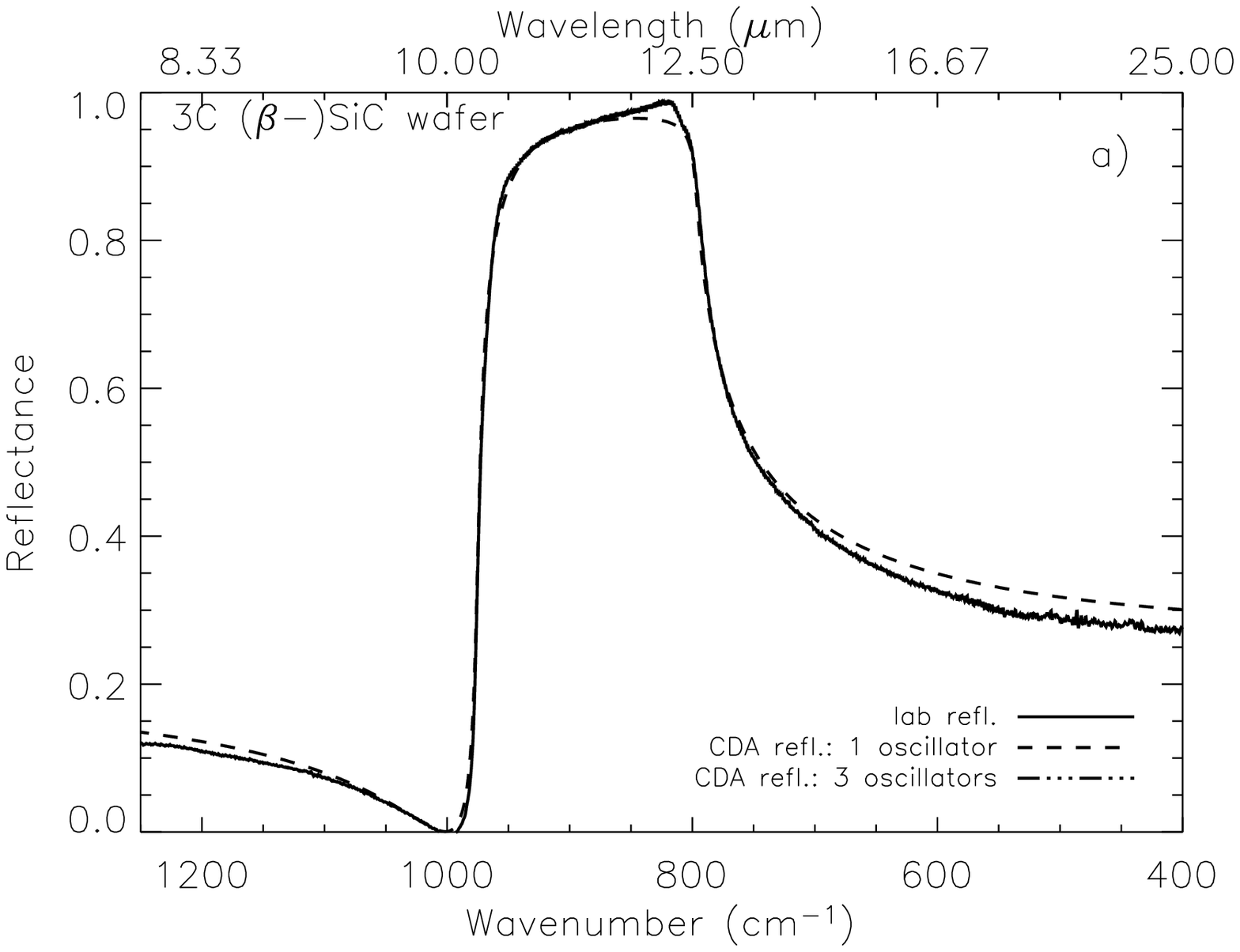}
    \end{minipage}\\
    \begin{minipage}[b]{0.5\textwidth}
       \includegraphics[width=\textwidth]{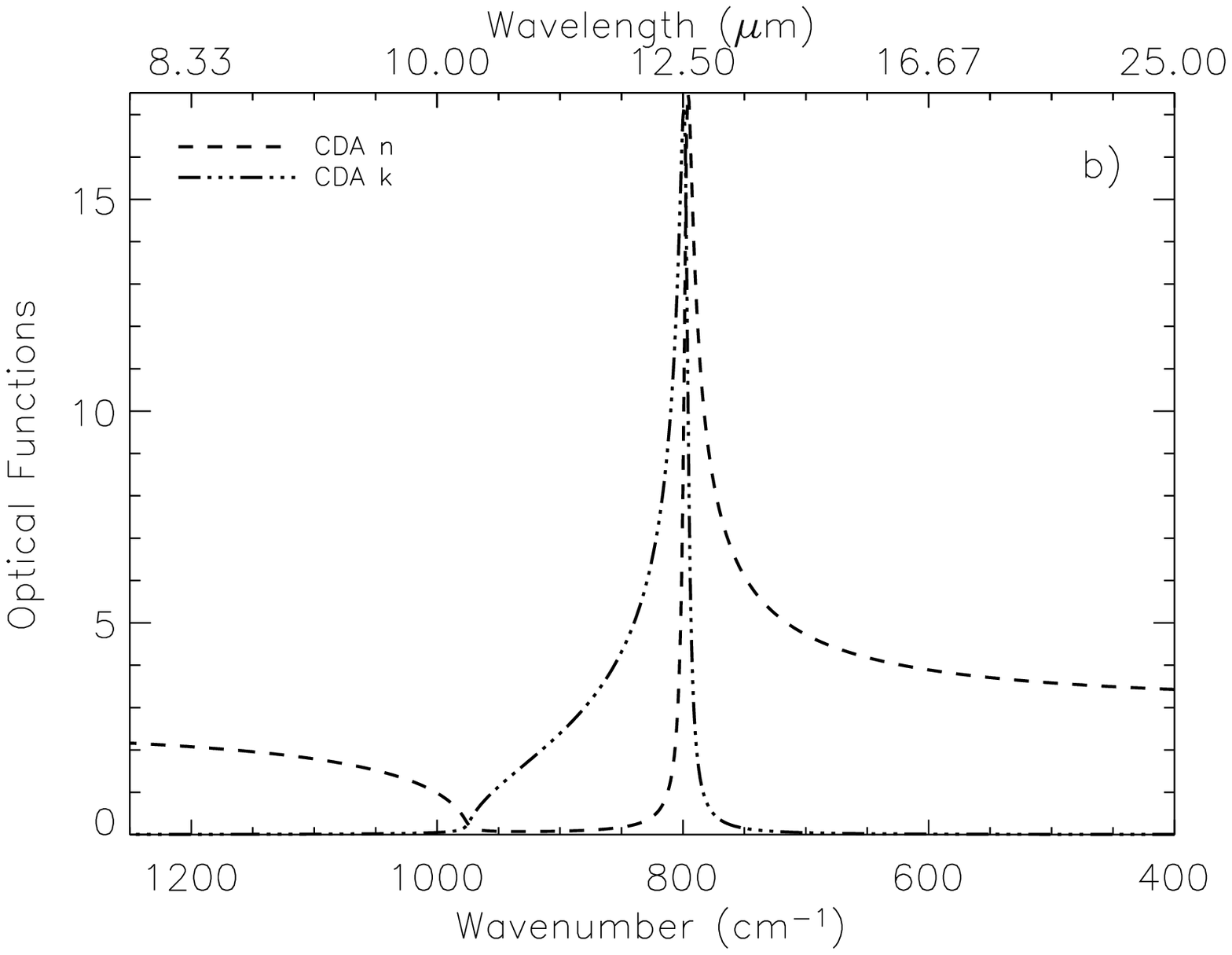}
    \end{minipage}
    \caption{Reflectivity at near-normal incidence of $\beta$-SiC 
  and derived $n$, $k$.
 (a) Laboratory reflectivity spectrum (solid line, scaled to 98\% 
 maximum reflectance), calculated reflectivity for 1 oscillator 
 (long dashed line: $\nu$~$=$~797.5~cm$^{-1}$, FWHM~$=$~6.0~cm$^{-1}$,
  $f$~$=$~3.5) and 3 oscillators (short dashed line: 
  $nu_{1}$~$=$~797.5 cm$^{-1}$, FWHM$_{1}$~$=$~2.66 cm$^{-1}$, 
  $f_{1}$~$=$~2.7, $nu_{2}$~$=$~802 cm$^{-1}$, FWHM$_{2}$~$=$~7.0 cm$^{-1}$, 
  $f_{2}$~$=$~0.68, $nu_{3}$~$=$~875 cm$^{-1}$, FWHM$_{3}$~$=$~7.0 cm$^{-1}$, 
  $f_{3}$~$=$~0.03).  An LO mode occurs at $\nu$~$=$~975~cm$^{-1}$. 
  (b) $n$ (solid line) and $k$ (dashed line) for the 1 oscillator fit.}
  \label{betawafer98}
 \end{figure}

We fitted our laboratory reflectance data to various numbers of 
oscillators via classical dispersion analysis to determine if
the derived optical constants and absorption coefficients are strongly
affected by both weak features predicted from symmetry analyses and by spectral artifacts 
induced by layering or small sample size.
Figures~\ref{alphaEparc1osc}-\ref{betawafer98} show the resulting
optical functions and the corresponding fitting parameters.
  By comparing and contrasting
these fits, we determine whether multiple oscillators or a single
oscillator best represents the behaviour of $\alpha$- and $\beta$-SiC.
 The $n$ and $k$ values obtained from single-crystal spectra
samples are not dependent on grain size.

For $\beta$-SiC, a classical dispersion analysis fit using one
oscillator matches the peak sites and the sloping top of the main 
reflectance peak well, 
but to fit the corner dip and the slight 
sway on the top of the peak, three oscillators are required 
(Figs.~\ref{betawafer98}, \ref{nkcomp}a).   
The LO mode is at $\nu$~$=$ 973~cm$^{-1}$.
The large breadth and the position of the 875~cm$^{-1}$ feature is 
consistent with assignment as an acoustic overtone.  Its presence 
is within the uncertainty of the measurements.  From Fig.~\ref{nkcomp}a,
$n$ and $k$ are slightly affected by the number of oscillators used.  
Using three oscillators instead of one makes the peak in $k$ more 
narrow and causes the maximum $k$ value to increase by 25\%. 
The total area under the peak for $k$ is not affected by the 
number of oscillators.  Similarly, $n$ differs little when the 
number of oscillators is varied.  Because no extra oscillators 
are expected at 802~cm$^{-1}$ for $\beta$-SiC,
we suspect the appearance of a peak at that location is associated 
with back reflections for this very thin wafer.  For the purpose of 
modelling the spectrum, the single oscillator fit 
($\nu$~$=$~797.5~cm$^{-1}$, FWHM~$=$~6.0~cm$^{-1}$, and 
$f$~$=$~3.5 with $\nu_{LO}$~$=$~975~cm$^{-1}$) suffices and 
agrees with symmetry analysis.

Fitting the moissanite $\vec{E} \bot \vec{c}$ reflectance spectrum
with one
oscillator shows that there is structure on the main band near 
both 800 and 950~cm$^{-1}$ (Fig.~\ref{moissEperp1osc}).  Also, the 
divot makes it difficult to constrain the TO position.  A one 
oscillator fit with peak wavenumber 
at 797.5~cm$^{-1}$, FWHM~$=$~4.5 cm$^{-1}$, and $f$~$=$~3.45 
is equally good (not shown).  Two additional oscillators are needed to characterize the structure near 800~cm$^{-1}$, and one additional oscillator is needed to account for the structure near 970~cm$^{-1}$ 
(Fig.~\ref{moissEperp1osc}).  
The best fit for two oscillators is $\nu_{1}$~$=$~797.5~cm$^{-1}$, 
FWHM$_{1}$~$=$~4.1~cm$^{-1}$, $f_{1}$~$=$~3.45 and 
$\nu_{2}$~$=$~970.0~cm$^{-1}$, FWHM$_{2}$~$=$~11.0~cm$^{-1}$, 
and $f_{2}$~$=$~0.0010.
The presence of a peak at 970~cm$^{-1}$ 
is in agreement with zone folding and Raman data.  Addition 
of the 970~cm$^{-1}$ makes very little difference to $n$ and $k$ 
(Fig.~\ref{nkcomp}).

For the 6H gray $\alpha$-SiC sample ($\vec{E} \bot \vec{c}$), a one
oscillator fit does not match the slope at high frequency 
(Fig.~\ref{alphaEperpc1osc}).  $n$ and $k$ differ very little between the fits, though multiple oscillators appear to make $k$ more narrow and peak at a higher value (Fig.~\ref{nkcomp}).
The main peak parameters differ very little 
among various fits to different numbers of oscillators for this 
reflectance spectrum.
The green and yellow $\alpha$-SiC samples ($\vec{E} \bot \vec{c}$) 
were fitted with a single oscillator (not shown).  For the green $\alpha$-SiC, the best fitting parameters are a peak position of 797.5~cm$^{-1}$, 
FWHM~$=$~6.0~cm$^{-1}$ and $f$~$=$~3.3; n.b., this does not fit the 
slope at high frequency well.  For the yellow $\alpha$-SiC, the best 
fitting parameters are a peak position at 798.0~cm$^{-1}$, 
FWHM~$=$~5.5~cm$^{-1}$ and $f$~$=$~3.5.  This TO position is not well 
constrained due to the presence of the divot.

Regarding possible spectral artifacts,
Spitzer et al. (1959b) also observed a divot in their reflectance spectrum 
for a polished SiC surface and a slope for an oxidized surface; 
they obtained their best data for a grown surface.
Data on $\alpha$-SiC, $\vec{E} \bot \vec{c}$ from figure 9.6 in 
Bohren \& Huffman (1983) resemble the results of Spitzer et al. (1959b) 
for the grown surface.  No divot exists, but the slope is greater than 
we observed for moissanite.
We note that the green and 6H gray $\alpha$-SiC samples both have thin 
layers perpendicular to $\vec{c}$ and high slopes at high $\nu$.
The 6H gray $\alpha$-SiC sample has a slightly stronger slope, but 
not much 
(Fig.~\ref{allEperpc}).  The yellow $\alpha$-SiC and moissanite samples
lack the slope but have the divot at low $\nu$.  These samples are 
single crystals; however, the yellow $\alpha$-SiC is the thinnest 
sample and its spectrum has the deepest divot.  The moissanite is not 
thin but has growth sectors (on the order of 0.5~mm thick).  
We conclude that the divot is connected with back reflections, not 
surface polish, because the green and yellow $\alpha$-SiC and moissanite 
samples are all polished by the manufacturers and have smooth surfaces.  
Given how difficult this divot was to fit via classical dispersion 
analysis, it cannot be a vibrational mode and is probably associated 
with a physical optics effect and the high reflectivity of the sample.  
We also conclude that the slope at high frequency 
is associated with the thin layers and is also a problem of physical 
optics.  Neither the slope nor the divot are intrinsic to the samples.  
Samples larger than 5~mm in diameter without growth sectors are needed to 
provide the best possible data for SiC.

For moissanite ($\vec{E} \| \vec{c}$), the main reflectance peak has a very large divot, which is an artifact, as in the $\vec{E} \bot \vec{c}$ polarization 
(Fig.~\ref{moissEparc1osc}). 
Because of the divot, the TO peak position is difficult to constrain.  
The best classical dispersion fit provides a high value of $\nu$ 
(802~cm$^{-1}$) that is uncertain.  The data do not reveal the presence of 
any of the folded modes or of acoustic overtone/combination bands.
The 6H gray ($\vec{E} \| \vec{c}$) sample can be fitted with one
oscillator at $\nu$~$=$~787.8~cm$^{-1}$ with a FWHM~$=$~5.5~cm$^{-1}$ 
and $f$~$=$~4.45 
(Fig.~\ref{alphaEparc1osc}).  
The peak position is hard to constrain, but clearly it is lower than the TO peak
for $\vec{E} \bot \vec{c}$.
The corresponding LO position is 
at $\nu_{LO}$~$=$~966.9~cm$^{-1}$.  The divot at low 
frequency is different,
being comprised of one broad mode
near 830~cm$^{-1}$ and having reduced intensity near the TO position, like 
the other samples (Fig.~\ref{allEparc}).
The latter fact should not impact the TO position of the main peak.
We attempted fits with various numbers of peaks.  The 4 oscillator fit in 
Fig.~\ref{alphaEparc1osc} confirms the low frequency TO position near 
787~cm$^{-1}$ for 6H gray $\alpha$-SiC ($\vec{E} \| \vec{c}$).  
The presence of two modes near 888~cm$^{-1}$ are consistent with zone 
folding and possible involvement of resonance with the acoustic modes.  
However, the narrow widths for these bands indicate that these are 
fundamentals, not overtones.  The broad feature near 835~cm$^{-1}$ 
appears to be a folded mode; its position is consistent with Raman spectra, 
although its breadth suggests it is an overtone/combination band.  In 
contrast, the reduced intensity near the TO position was difficult to 
fit, which is the hallmark of a spectral artifact and consistent with the 
problems incurred in the other samples.  The strong layering in the 6H 
gray $\alpha$-SiC sample could contribute to the artifacts.  The fit
parameters that should represent the intrinsic behaviour of this 
particular sample is $\nu$~$=$~787.8~cm$^{-1}$ (FWHM~$=$~5.5 cm$^{-1}$, 
$f$~$=$~4.35), 836.0~cm$^{-1}$ (FWHM~$=$~30.0~cm$^{-1}$, $f$~$=$~0.10), 
883.7~cm$^{-1}$ (FWHM~$=$~2.5~cm$^{-1}$, $f$~$=$~0.0041) and 888.5~cm$^{-1}$ 
(FWHM~$=$~3.3~cm$^{-1}$, $f$~$=$~0.0033).  The intense LO mode occurs at 
968.9~cm$^{-1}$, consistent with Raman measurements.  We suggest that the 
sample has impurities 
connected with the strength of the folded modes.  These weak modes add 
structure to the $n$ and $k$ spectra but do not change the peak maximum.

\subsection{Calculated absorbance}

\begin{figure}
     \centering
    \includegraphics[width=8cm]{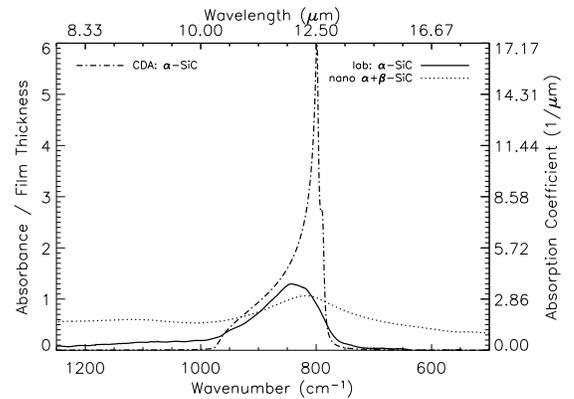}
    \caption{Laboratory thin film absorbances (Speck et al. 1999; 
    Speck et al. 2005) divided by estimated 
    sample thicknesses ($d$~$\sim$~0.5~$\mu$m) compared to calculated 
    classical dispersion analysis absorbance coefficients for $\alpha$-SiC.  
    Calculated absorbance coefficients for the remaining $\alpha$-SiC 
    samples are indistinguishable.}
    \label{alphlababs}
   \end{figure}

\begin{figure}
     \centering
    \includegraphics[width=8cm]{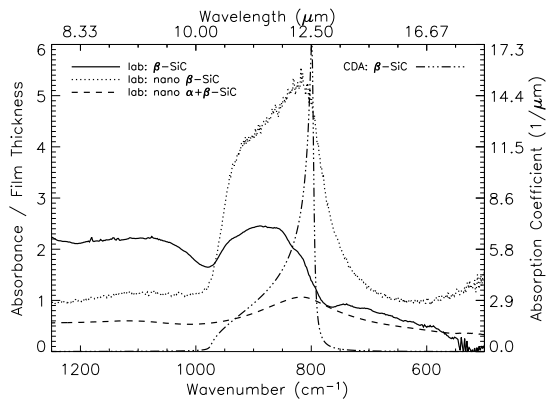}
    \caption{Laboratory thin film absorbances (Speck et al. 1999; 
    Speck et al. 2005) divided by estimated
    sample thicknesses ($d$~$\sim$~0.15~$\mu$m for $\beta$-SiC, 
    0.5~$\mu$m for nano samples) compared to calculated 
    classical dispersion analysis
    absorbance coefficients for a 1 oscillator fit to $\beta$-SiC 
    (Fig.~\ref{betawafer98}).  Given the uncertainty in 
    estimated sample thicknesses, laboratory values 
    can be scaled by factors of 3-5.}
    \label{betalababs}
    \end{figure}

\begin{figure}
    \centering
    \includegraphics[width=8cm]{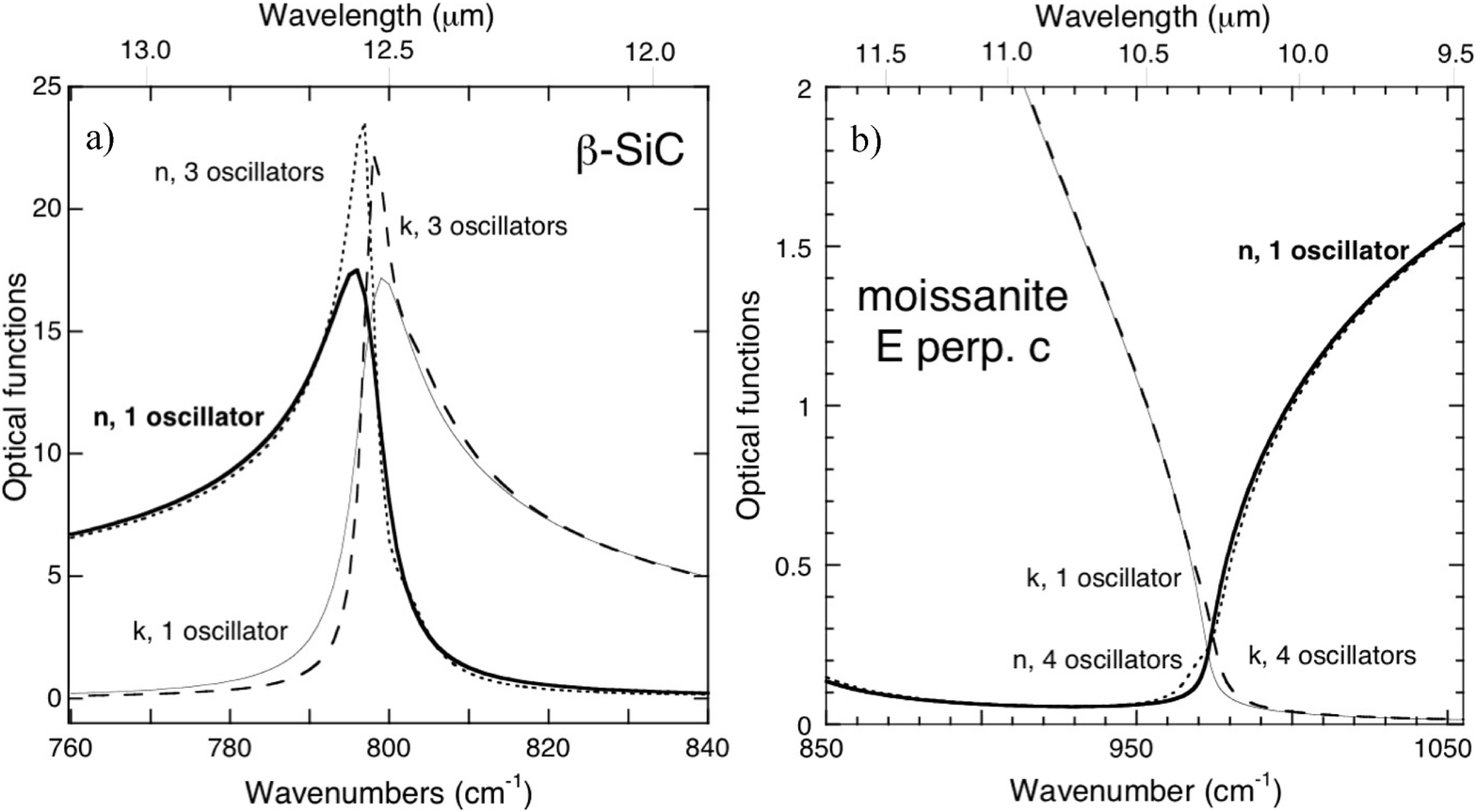}
    \caption{Comparison of $n$ and $k$ values calculated via 
    classical dispersion for single and multiple oscillator fits: 
    (a) $\beta$-SiC (Fig.~\ref{betawafer98}, \S~4.2), and 
    (b) $\alpha$-SiC (moissanite $\vec{E} \bot \vec{c}$, 
    Figs.~\ref{moissEperp1osc}b).  In (a), peak heights for 
    $n$ and $k$ increase if more oscillators are used in the fit.
    In (b), adding oscillators to account for structure at
    $\nu$~$\sim$~800, 970~cm$^{-1}$ improves the fit to the 
    laboratory reflectance spectrum, but the difference to $n$ and
    $k$ is small.}
    \label{nkcomp}
  \end{figure}

Absorbances calculated via classical dispersion analyses  
are cross-checked through comparison with measured absorbance of the 
$\alpha$- and $\beta$-SiC samples (Fig.~\ref{alphlababs},\ref{betalababs}).
In contrast to the calculated $A(\nu)$, laboratory absorbances ($a_{SCM}$) 
are dependent on grain size via

\begin{equation}
a_{SCM} = \frac{A(\nu)d}{2.3026} - 2\log_{10}(1 - R).
\label{lababseq}
\end{equation}

$a_{SCM}$ is converted for direct comparison to $A(\nu)$ in the figures.  
For $\alpha$-SiC, 
a weighted average,  

\begin{equation}
a_{avg} = (2a_{\bot} + a_{\|})/3,
\label{avgabseq}
\end{equation}

\noindent
is used in the comparison of Fig.~\ref{alphlababs}. Averaging provides 
one main peak and a shoulder for $\alpha$-SiC.  The position of the main 
peak matches that of $\beta$-SiC (Fig.~\ref{betalababs}), but the 
shoulder is at lower frequency.  Real, measured spectra have 
strongly rounded peaks due to light leakage 
between cracks in thin-film samples or around the particulates in 
dispersions.  In measured spectra, the $\alpha$-SiC peak 
occurs at a lower frequency than for $\beta$-SiC, consistent 
with our calculations.  Because of the light leakage, the two 
peaks for $\alpha$-SiC are blended into one band.  In 
Fig.~\ref{alphlababs}, the measured spectrum of 
$\alpha$-SiC fits into the envelope.  Some of the differences 
between the calculated and measured absorbances are due 
to reflectivity not being subtracted from the raw data; in 
these diamond anvil cell measurements, the spectral baseline 
involves both the reflectivity of the diamonds and of the sample, 
and it is not clear how best to proceed with the subtraction. 
In dispersions, the spectral baselines are uncontrolled, consisting 
of sample and medium properties.  For $\beta$-SiC, the measured 
peaks are much more rounded than in the calculations, and more 
rounded than the peaks measured for $\alpha$-SiC   
(cf. Fig.~\ref{alphlababs},\ref{betalababs}).  
Because the $\beta$-SiC chips are harder, it is more difficult 
to make a thin film; this results in a more rounded profile.  
The nano-$\beta$-SiC samples are softer and form a better thin film, 
producing a peak very close to the ideal, and also a large shoulder 
at the LO position.  The prominence of the LO mode is expected for 
the $\beta$-SiC absorbance spectrum which consists of a doubly 
degenerate TO mode and a single LO mode.  Hence absorbance peaks 
round and shift to high frequency in measurements as compared to 
classical dispersion analysis calculations 
(Fig.~\ref{betalababs}).  For the hexagonal samples, symmetry suggests 
something much different: a degenerate planar optic mode near the 
cubic TO position and 1 axial optic mode near the LO position.  That is, 
the TO and LO modes that are strongly coupled in the cubic structure are 
decoupled in the hexagonal structure.  
For $\vec{E} \bot \vec{c}$ in $\alpha$-SiC, the TO and LO modes 
are like those in $\beta$-SiC.  The behaviour of the absorbance spectrum 
for $\alpha$-SiC ($\vec{E} \| \vec{c}$) is confusing because several 
criteria must be met that seem to be contradictory.  One is that the 
TO mode must be lower in frequency than the LO mode, despite the fact that 
symmetry suggests a main mode occurs near $\nu$~$=$~970~cm$^{-1}$ for the 
axial configuration.  Another is that the same dipoles exist, so the 
stretching frequency cannot be much different.  The end result is that 
the LO position is preserved and  the TO position is derived from a 
zone-folded mode.  It seems that more strength is shifted to the TO 
position in the process.  But it is also expected that the hexagonal 
particles are preferentially oriented and that the measured absorption 
is skewed to the TO position.  These effects combined indicate that 
less energy goes toward the LO position in $\alpha$-SiC than in 
$\beta$-SiC, consistent with the rounding of the peaks and the lower 
frequency of the $\alpha$-SiC peak maximum.

%

%

\section{Discussion}

Past studies are divided on whether 
the crystal structure of SiC can be determined from IR spectra (in favor: 
Borghesi et al. 1985, Speck et al. 1999; opposed: Spitzer et al. 1959a, b, 
Papoular et al. 1998, Andersen et al. 1999a,b, Mutschke et al. 1999).  
\S~4 shows that spectroscopic differences exist
between certain orientations of $\alpha$- and $\beta$-SiC.
This section compares $\beta$-SiC to $\alpha$-SiC from 
the laboratory perspective and discusses implications for astronomers.

\subsection{Comparisons to previous works}

Spitzer et al. (1959a) obtained reflectivity data from a thin (0.06~$\mu$m) 
film of $\beta$-SiC that was vapour-deposited on a Si surface 
($\nu_{TO} =$~793.6~cm$^{-1}$, $\Gamma =$~8.4~cm$^{-1}$, $f = 3.30$).  
The film was slightly irregular in appearance and 
thin enough to transmit light at all frequencies (their fig.~3).
Because the TO mode dominates those
spectra, the TO peak position should be as accurate as their
spectrometer could provide.  However, the width of 
8.5 cm$^{-1}$ is large compared to that from 6H crystals,
which probably stems from the LO mode being disproportionately 
large in the thin-film spectra.  This effect occurs commonly due 
to non-normal beam incidence, wedging of the film, or irregular 
film thickness (Berreman 1963).  The LO value obtained from the 
parameters of Spitzer et al. (1959a) will be influenced by the 
large $\Gamma$ value.

The $\alpha$-SiC $\vec{E} \bot \vec{c}$ data 
shown in fig.~9.6 of 
Bohren \& Huffman (1983) is historically important because it was used in 
the widely cited work by Laor \& Draine (1993). 
Because Bohren \& Huffman (1983) only provide reflectance data for 
$\vec{E} \bot \vec{c}$ and did not provide experimental details, such as which type of 
$\alpha$-SiC was used, direct comparison 
between our 6H-SiC data and past studies is weighted more 
heavily toward results of the Spitzer et al. (1959b) study, which 
provides data for both polarizations of 6H single-crystals.

Our results suggest that the $n(\lambda)$ and $k(\lambda)$ for
$\nu$~$=$~797.5~cm$^{-1}$, 
FWHM~$=$~4.0--6.0~cm$^{-1}$ and $f$~$=$~3.3--3.45 best represents 
$\alpha$-SiC $\vec{E} \bot \vec{c}$.  These values are quite similar 
to the $\vec{E} \bot \vec{c}$ results of Spitzer et al. 
(1959b) for 6H-SiC ($\nu_{TO}$~$=$~793.9~cm$^{-1}$, 
FWHM~$=$~4.8~cm$^{-1}$, $f$~$=$~3.30) 
and Bohren \& Huffman (1983) for $\alpha$-SiC 
($\nu_{LO} =$~969.2~cm$^{-1}$, $\nu_{TO} =$~793~cm$^{-1}$, 
$\Gamma =$~4.7~cm$^{-1}$).  
Thus, we confirm past findings that $f$ is high and that the FWHM of the main 
SiC peak is of the order of 5 cm$^{-1}$.  For 
dispersive instruments such as the ones 
used by Spitzer et al. (1959b) and presumably Bohren \& Huffman (1983), resolutions depend on the 
slits used and vary with wavelength. Low resolution is suggested by the spacing of data points
in previous figures by these authors. High 
resolution is needed to properly depict profiles of sharp, steeply rising peaks
(cf. Bowey et al. 2001), such as the reflectance peaks for SiC.  The few 
cm$^{-1}$ difference for the reported TO positions may either be due to resolution difference or to older, dispersive instruments lacking 
internal calibration that exists in modern FT-IR spectrometers.  
Peak positions obtained from our high resolution reflectance 
measurements agree closely with Raman studies (Table~1); 
because Raman peaks are 
narrower than IR reflectance or absorbance peaks and have FWHM 
values similar to those of the TO modes in the dielectric functions, 
the TO values presented in this study can be considered more accurate 
than the lower values previously reported.  Spectral artifacts observed 
by Spitzer et al. (1959b) exist for our samples as well and are 
consistent with the presence of back reflections from internal surfaces.  This 
deduction is based on the different thicknesses of the internal reflection 
surfaces in our various samples.  In contrast to the Bohren \& 
Huffman (1983) $\alpha$-SiC data and Spitzer et al. (1959b) data 
for the 6H-SiC grown surface, we see a new divot in our 6H-SiC spectra.  
The slope of the spectral profiles in past
studies is greater than what we observed for moissanite.  The weak 
oscillator seen in our 6H-SiC data at 
$\nu$~$=$~970~cm$^{-1}$, FWHM~$=$~11.0~cm$^{-1}$ and 
$f$~$=$~0.001 is real, but its presence barely alters $n$ and $k$ 
and will not affect RT models.  

For $\vec{E} \| \vec{c}$, Spitzer et al. (1959b) observed one weak band near 
883~cm$^{-1}$.  We observed a doublet due to use of higher resolution.  
Their peak parameters of $\nu$~$=$~785.9~cm$^{-1}$, 
FWHM~$=$~5.5~cm$^{-1}$, and $f$~$=$~3.3 for a single oscillator fit 
(with a LO mode at 966.9~cm$^{-1}$) are quite similar to ours and 
within the experimental uncertainty, given the results for various 
samples.  Their TO positions for both polarizations are low, consistent 
with the resolution or calibration.  We conclude that the best 
representation of $\vec{E} \| \vec{c}$ for essentially pure $\alpha$-SiC is 
the 1 oscillator fit for our 6H gray sample. 
 
There exists precedent in the SiC literature to average spectral parameters 
from many laboratory studies to obtain 
TO frequency positions, LO frequency positions or oscillator 
strengths, and FWHM values for SiC.  For example, 
Mutschke et al. (1999) give the following frequencies, averaged 
from experimental studies from the 1960's to 1990's (their table 1): 
(for 3C-SiC) $E_{1T} = A_{1T} =$ 795.9 cm$^{-1}$, 
$E_{1L} = A_{1L} =$ 972.3 cm$^{-1}$, and (for 6H-SiC) 
$E_{1T} =$ 797.0 cm$^{-1}$, $A_{1T} =$ 788.1 cm$^{-1}$, 
$E_{1L} =$ 969.9 cm$^{-1}$, $A_{1L} =$ 965.3 cm$^{-1}$.  
The frequency positions presented in our paper seem at first glance to 
be farther from values based on Raman measurements
(from Hofmann et al. 1994, 3C-SiC: $E_{1T} = A_{1T} =$ 795.7 cm$^{-1}$, 
$E_{1L} = A_{1L} =$ 979.0 cm$^{-1}$; 
6H-SiC: $E_{1T} =$ 797.0 cm$^{-1}$, $A_{1T} =$ 788.1 cm$^{-1}$,
$E_{1L} =$ 969.9 cm$^{-1}$, $A_{1L} =$ 965.3 cm$^{-1}$); however, if 
the data by Spitzer et al. (1959a,b) had been included in the 
averages, the mean values reported by Mutschke et al. (1999) would 
be shifted by an amount up to 0.4 cm$^{-1}$.

We have discussed our data in context with some of the major past 
laboratory studies and compilations of experimental data on SiC.  
To determine which spectral parameters are 
most representative of SiC, we strongly encourage readers to assess 
individual datasets on a case-by-case basis.  This is because
a single set of extremely precise data may be as accurate or more 
so than values obtained by averaging several sets of moderately precise data 
(McKenna \& Hodges 1988; Kohn \& Spear 1991).  Averaging datasets also 
presumes that all errors in each dataset are random.  
If, instead, systematic errors arise either in data collection or analysis, 
then averaging will not provide the true values (e.g., Bevington 1969).  
Systematic errors exist in previously published SiC reflectance data, 
e.g., the differences between the instrumental 
resolutions of dispersive and FT-IR spectrometers can contribute up to 
factors of a few cm$^{-1}$ difference in the reported positions of peaks.  
For these reasons, we discourage readers 
from directly combining our (or any FT-IR) SiC data with older results. 

\subsection{Differences between $\beta$- and $\alpha$-SiC; implications for 2H}

By obtaining reflectivity spectra from different samples and polytypes,
 we have constrained the peak parameters of the main SiC features and the 
 optical functions below UV frequencies where metal-anion charge transfer 
 exists.  
Peak parameters for $\beta$ and $\vec{E} \bot \vec{c}$ of 6H $\alpha$ SiC
were virtually indistinguishable 
($\nu$~$=$~797.5~cm$^{-1}$, FWHM~$=$~5--6~cm$^{-1}$, $f$~$\sim$~3.5).  
The parameters for $\vec{E} \| \vec{c}$ of 6H $\alpha$-SiC were also 
similar but the peaks occurred at slightly lower frequencies.  This 
behaviour is consistent with symmetry analysis 
and Raman data (Table~1), indicating that peak positions from past
IR studies err by several cm$^{-1}$.
Additional weak modes exist (acoustic for $\beta$-SiC, zone-folded 
for 6H $\alpha$-SiC), but most of these are too weak and broad
to affect $n(\lambda)$ and $k(\lambda)$.  For samples with impurities 
(excess C), zone-folded modes near 880~cm$^{-1}$ are relatively 
strong and sharp.  The presence of these modes in $\vec{E} \| \vec{c}$ 
alters $n(\lambda)$ and $k(\lambda)$ in a minor way by shifting 
the main peak $\nu$ down by $<$ 2 cm$^{-1}$.
Thus, we believe that the 1 oscillator fit should be used for 
$\beta$-SiC or for ``pure'' SiC.  The doublet in
$\alpha$-SiC ($\vec{E} \| \vec{c}$) is real and should be accounted for; 
we recommend use of a 3 oscillator fit for that 
or for SiC with carbon excess or stacking anomalies.

The peak parameters for 6H $\alpha$-SiC can also be used for 
2H $\alpha$-SiC.  The symmetry analysis and Raman data 
(\S~2) indicate that the IR peak parameters for 
$\vec{E} \bot \vec{c}$ of 2H should be similar to those for 6H, 
i.e., our data on moissanite represent 2H for $\vec{E} \| \vec{c}$.
For $\vec{E} \| \vec{c}$ of 2H $\alpha$-SiC, a weak mode should 
occur near 838 cm$^{-1}$ and 
the strong mode is downshifted in frequency to about 965 cm$^{-1}$.  
Although folded modes are not expected near 888 cm$^{-1}$, a broad 
feature is seen in this region for 3C SiC.  We suggest that 
2H $\alpha$-SiC ($\vec{E} \| \vec{c}$)
is represented by the parameters found for 
the main peak and 838 cm$^{-1}$ mode of 6H SiC, with the 
broad oscillator near 881 cm$^{-1}$ from $\beta$-SiC.

\subsection{Recommendations for radiative transfer modelers}

The purpose of this paper is to provide accurate
optical data for the polytype of SiC most 
commonly found in astronomical environments ($\beta$-SiC) and the most 
commonly manufactured SiC polytype ($\alpha$-SiC), so that we and others 
may construct improved radiative transfer (RT) models.  
The advantages to modelers in using this dataset are that (1) we 
measured $\beta$-SiC and $\alpha$-SiC in all orientations in the 
same laboratory, (2) we have used the full classical dispersion
equations 
to arrive at $n(\lambda)$ and $k(\lambda)$, and (3) our $n$ and $k$ 
data are not dependent on grain size. 
The electronic files for $n$ and $k$ (Tables~3--7, available at the 
CDS, where Column~1 is $\lambda$ in $\mu$m, Column~2 is $n$, and 
Column~3 is $k$) have been prepared assuming that 
these will be inputs to the RT code DUSTY (Nenkova et al. 2000).
The original $n$ and $k$ data files we generated 
contained over 4000 points; we have regridded these 
data to a smaller number of points as required for DUSTY, maintaining high 
$\lambda$ resolution where needed (i.e., where the slope of the 
data is steep) and using low $\lambda$ resolution where the slope is 
shallow or flat.  We provide the regridded $n$ and $k$ data 
electronically; readers may contact the authors 
for the 4000 point high resolution files.   
If using DUSTY, the default wavelength grid is too coarse and 
must be altered to provide sufficient spectral resolution at 10--13~$\mu$m 
where SiC has its strong resonance feature.  
The spectral resolution in this region is 0.005~$\mu$m.  
DUSTY and many other RT models invoke Mie theory 
to calculate absorption and scattering cross-sections when a user 
supplies $n$ and $k$. This implicitly assumes spherical grains. As 
discussed below, spherical grains are probably inappropriate in 
most astrophysical environments.
Thus, we recommend supplying 
absorption and scattering coefficients for non-spherical grains calculated 
from our $n$ and $k$ instead of directly supplying 
$n$ and $k$ to DUSTY. 

The intrinsic 
shape for circumstellar SiC grains remains unknown.  Some recent works 
have modelled SiC grains as spheres 
(cf., Gauba \& Parthasarathy 2004; Thompson et al. 2006; 
Lunttila \& Juvela 2007).  Jiang et al. (2005) argue that spheres are adequate 
for $\mu$m- or sub-$\mu$m-sized SiC grains at 
$\lambda >$ 11~$\mu$m because the 
broadening of the 11.3 and 21~$\mu$m features (expected if one assumes 
that the grains are a continuous distribution of ellipsoids) is not 
likely to differ in the Rayleigh regime.
Rayleigh scattering suffices when particles are small.
However, the strength of the absorption as well as particle size 
is required to ascertain which scattering regime pertains (Lynch \& 
Mazuk, 1999).  Because SiC has high absorption and nearly perfect 
reflection at peak center and the SiC particle sizes expected in 
astronomical environments are close to the sizes where Rayleigh 
scattering no longer applies, other shapes should be considered.
It is clear that distributions of more complex shapes (e.g., 
Continuous Distribution of 
Ellipsoids, CDE, Bohren \& Huffman 1983; Distribution of Hollow Spheres, 
Min et al. 2003; aggregates, Andersen et al. 2006 and references therein) 
should be used in rigorous parameter space explorations of SiC grains.
Most of the standard non-spherical grain shape distributions 
give rise to a feature at the relevant wavelength and with the 
broader observed feature width.
  
There is no clear consensus on what the grain size distribution 
for SiC grains should be (see review by Speck et al. 2008, submitted).  
SiC grains should deviate from the MRN grain 
size distribution (i.e., $m_{d}(a)$ $\propto$ $a^{-q}$, 
where $m_{d}$ is the number of grains in the size interval 
($a, a + da$) and $q = 3.5$ 
$a_{min}$~$=$~0.005~$\mu$m and $a_{max}$~$=$~0.25~$\mu$m; 
Mathis et al 1977), used in works cited in 
\S~1 and the default grain size distribution in DUSTY.  
The MRN distribution is based on UV and visible data and gives no 
information for IR wavelengths.  In addition, SiC dust is generally found in 
circumstellar, not interstellar, dust; circumstellar dust is known not 
to be MRN in size or composition.  Some 
previously published $n$ and $k$ data 
are based on specific samples and are only relevant to the
grain sizes used in the lab experiments which provided the raw data. In
both P\'{e}gouri\'{e} (1988) and Laor \& Draine (1993), the
data originates from SiC-600 (ground and sedimented) sample from Borghesi
et al. (1985). The Borghesi et al. (1985) sample is 
99.3\% pure 6H $\alpha$-SiC, with a typical grain size of 0.04$\mu$m
and a large grain size tail 
which goes as $\propto a^{-2.1}$  ($a_{max} \sim$~0.5~$\mu$m).
From the IR spectrum, one infers that
the grains are ellipsoidal
rather than spherical or irregular (even though the unground grains are
clearly irregular in morphology). 
$n$ and $k$ derived from the Borghesi et al. (1985) data should 
thus only be applied to the grains in a similar grain size/shape regime.
The $n$ and $k$ values presented here offer an advantage over past 
datasets in that the new values may be used for extended grain size 
distributions such as MRN or KMH (Kim, Martin \& Hendry 1994).

Several recent papers have discussed the occurrence of a mid-IR 
absorption feature in the spectra of a few carbon stars (e.g., 
Cl\'{e}ment et al. 2003, 2005; Speck et al. 2005, 2006, 2008, 
submitted; Pitman et al. 2006).  Whilst this feature has been
attributed to SiC, it tends to occur at a shorter wavelength.  
Investigating grain size and shape effects with the new optical 
constants may allow us not only to fit this mid-IR feature, but 
also to determine what physical parameters differ in the stars that 
exhibit this absorption feature.  In addition, past
studies of grain size and shape based on the optical constants of 
P\'{e}gouri\'{e} (1988) and Laor \& Draine (1993) 
should be revisited.

\section{Conclusions}

\begin{enumerate}
\item{
Previously reported laboratory reflectivity data are for 
$\alpha$-SiC, whereas SiC dust surrounding astronomical objects is $\beta$-SiC;
the optical functions $n(\lambda)$ and $k(\lambda)$ for 
these differ slightly.
We have provided electronic table values for optical functions 
$n(\lambda)$ and $k(\lambda)$ at $\lambda =$~0.05--2000~$\mu$m derived 
via classical dispersion analysis from room temperature, 
near-normal incidence laboratory specular reflectance spectra of 
several samples of both 3C ($\beta$-) and 6H ($\alpha$-) SiC.  These $n$ and 
$k$ values are 
independent of grain size, do not change with scaling of 
the raw reflectance spectra, and may be used when modelling optically 
thin conditions.
\item{We have investigated whether 
fitting SiC IR spectra with one or more oscillators makes a difference 
to $n$ and $k$.  We suggest that a one oscillator fit best 
represents pure SiC.  A three oscillator fit best represents impure SiC 
(due to carbon excess or layer stacking).  
$\beta$-SiC and $\alpha$-SiC $\vec{E} \bot \vec{c}$ are best fitted 
using one oscillator with peak parameters $\nu$~$=$~797.5~cm$^{-1}$, 
FWHM~$=$~4.0--6.0~cm$^{-1}$, and $f$~$=$~3.5.  Adding a peak at 
$\nu =$~970~cm$^{-1}$ makes little difference to $n$ and $k$.  
For pure $\alpha$-SiC $\vec{E} \| \vec{c}$, 1 oscillator works 
well; as impurities increase, 3 oscillators are needed.
Strong layering of the sample contributes 
spectral artifacts to the reflectance spectra of $\alpha$-SiC 
when $\vec{E} \| \vec{c}$; thus, the fit 
parameters for that orientation carry more uncertainty.  The main 
difference between $n$ and $k$ for the different orientations of 
6H $\alpha$-SiC is that the TO peak position for $\vec{E} \| \vec{c}$ is 
lower than that of $\vec{E} \bot \vec{c}$.  Because the fit parameters for 
6H $\alpha$-SiC ($\vec{E} \bot \vec{c}$) are similar to those 
of $\beta$-SiC, this is also the 
main difference between $\alpha$-SiC and $\beta$-SiC.
Peak parameters for the moissanite sample may also be used for 
meteoritic 2H $\alpha$-SiC.} 
\item{
Peak positions in our SiC reflectance spectra are 
in agreement with 
Raman determinations of peak positions for 
SiC, which should be highly accurate due to the narrow widths, 
instrument calibrations, and large number of measurements made.
Our calculated peak widths and strengths for 6H-SiC
are in agreement with 
published Raman and reflectance studies.
Past data for $\beta$-SiC were collected from a thin film 
and give information on the TO mode, but the presence of the LO mode 
may affect those results.}
Differences in peak positions between this and past works are due 
to calibration, instrumental resolution, or a combination of both effects.
Apart from the fact that past
$n$ and $k$ datasets were compiled from several studies,  
older $n$ and $k$ values are not appropriate for dust that is optically thin 
at all frequencies, nor are these values appropriate for 
all grain sizes; thus, $n$ and $k$ data from past studies should be 
used carefully.  Comparing and contrasting results of 
more sophisticated radiative transfer models that use these 
new $n$ and $k$ values 
as inputs will allow the community to test competing hypotheses on 
grain size and shape effects in astronomical environments in future work.}

\end{enumerate}
			
\begin{acknowledgements}
     This work was supported by NASA APRA04--000--0041, NSF--AST 0607341, 
     and NSF--AST 0607418 and partly performed at the Jet Propulsion 
     Laboratory, California Institute of Technology, under contract to 
     the National Aeronautics and Space Administration.  KMP is supported 
     by an appointment to the NASA Postdoctoral Program, administered by 
     Oak Ridge Associated Universities.  The authors thank M. Meixner, 
     I. Parkin, G. C. Clayton, C. Dijkstra, and A. Koziol for their help 
     and H. Mutschke for careful reviews of 
     our manuscript leading to improvements in the presentation of 
     these results. 
\end{acknowledgements}

\end{document}